\documentclass[lettersize,journal]{IEEEtran}
\usepackage{amsmath,amsfonts}

\usepackage{cite}
\usepackage{amsmath,amssymb,amsfonts}
\usepackage{bm}
\usepackage[T1]{fontenc}

\usepackage{caption} % 确保已加载
\usepackage{titlesec}
\titlespacing{\section}{0pt}{1ex}{1ex}
\titlespacing{\subsection}{0pt}{1ex}{0.5ex}

\usepackage[linesnumbered,ruled,vlined]{algorithm2e}
\SetAlFnt{\small}
\SetAlCapFnt{\small}
\SetAlCapNameFnt{\small}
\let\oldnl\nl
\newcommand{\nonl}{\renewcommand{\nl}{\stepcounter{AlgoLine}\let\nl\oldnl}}

\usepackage{xcolor}
\usepackage{graphicx}
\usepackage{textcomp}

\usepackage{array}
\usepackage{tabularx}
\usepackage{float}
\usepackage{stfloats}

\usepackage{subfigure}
\usepackage[caption=false,font=normalsize,labelfont=sf,textfont=sf]{subfig}

\usepackage{url}
\usepackage{verbatim}
\usepackage{CJKutf8}

\SetKwInput{KwIn}{Input}
\SetKwInput{KwOut}{Output}
\SetKwInput{KwRet}{Return}
\DontPrintSemicolon  % 不自动添加分号
\usepackage{colortbl} % 用于单元格背景颜色
\definecolor{tablegray}{gray}{0.75} % 稍微深一点的灰色，如果lightgray太浅
\SetKwComment{tcp}{// }{}  % 使用 \, 来添加更小的间距
\SetCommentSty{mycommentstyle}

\usepackage{xr}
\makeatletter
\newcommand*{\addFileDependency}[1]{
  \typeout{(#1)}
  \@addtofilelist{#1}
  \IfFileExists{#1}{}{\typeout{No file #1.}}
}
\makeatother

\newcommand*{\myexternaldocument}[1]{
    \externaldocument{#1}
    \addFileDependency{#1.tex}
    \addFileDependency{#1.aux}
}
\myexternaldocument{Supplementary_materials}

\def\BibTeX{{\rm B\kern-.05em{\sc i\kern-.025em b}\kern-.08em
    T\kern-.1667em\lower.7ex\hbox{E}\kern-.125emX}}

\title{A segment anchoring-based balancing algorithm for agricultural multi-robot task allocation with energy constraints}

\author{
Peng Chen,
Jing Liang*,~\IEEEmembership{Senior Member,~IEEE},
Kang-Jia Qiao,
Hui Song,
Tian-lei Ma,~\IEEEmembership{Member,~IEEE},
Kun-Jie Yu,~\IEEEmembership{Member,~IEEE},
Cai-Tong Yue,~\IEEEmembership{Member,~IEEE},
Ponnuthurai Nagaratnam Suganthan,~\IEEEmembership{Fellow,~IEEE},
Witold Pedrycz,~\IEEEmembership{Life Fellow,~IEEE}
\thanks{This work was supported in part by the National Natural Science Foundation of China under Grant U23A20340 and in part by the National Key Research and Development Program of China under Grant 2022YFD2001200. \textit{(Correspoinding author: Jing Liang.)}  } 
\thanks{Peng Chen, Kang-Jia Qiao, Tian-lei Ma, Kun-Jie Yu, and Cai-Tong Yue are with the School of Electrical and Information Engineering, Zhengzhou University, Zhengzhou 450007, China.}
\thanks{Jing Liang is with the School of Electrical Engineering and Automation, Henan Institute of Technology, Xinxiang 453003, China, and also with the School of Electrical and Information Engineering, Zhengzhou University, Zhengzhou 450007, China (e-mail: liangjing@zzu.edu.cn).}
\thanks{Hui Song is with the School of Engineering, RMIT University, Melbourne, VIC, 3000, Australia. }
\thanks{Ponnuthurai Nagaratnam Suganthan is with the KINDI Center for Computing Research, College of Engineering, Qatar University, Doha, Qatar.}
\thanks{Witold Pedrycz is with the Department of Electrical and Computer Engineering, University of Alberta, Edmonton, Canada, and also with the Systems Research Institute of the Polish Academy of Sciences, Warsaw, Poland.}
}

\IEEEaftertitletext{\vspace{-2.5em}}

\begin{document}
\maketitle
\thispagestyle{empty}

 % Adjust the -3em value as needed

\begin{abstract}
Multi-robot systems have emerged as a key technology for addressing the efficiency and cost challenges in labor-intensive industries. In the representative scenario of smart farming, planning efficient harvesting schedules for a fleet of electric robots presents a highly challenging frontier problem. The complexity arises not only from the need to find Pareto-optimal solutions for the conflicting objectives of makespan and transportation cost, but also from the necessity to simultaneously manage payload constraints and finite battery capacity. When robot loads are dynamically updated during planned multi-trip operations, a mandatory recharge triggered by energy constraints introduces an unscheduled load reset. This interaction creates a complex cascading effect that disrupts the entire schedule and renders traditional optimization methods ineffective. To address this challenge, this paper proposes the segment anchoring-based balancing algorithm (SABA). The core of SABA lies in the organic combination of two synergistic mechanisms: the sequential anchoring and balancing mechanism, which leverages charging decisions as `anchors' to systematically reconstruct disrupted routes, while the proportional splitting-based rebalancing mechanism is responsible for the fine-grained balancing and tuning of the final solutions' makespans. Extensive comparative experiments, conducted on a real-world case study and a suite of benchmark instances, demonstrate that SABA comprehensively outperforms 6 state-of-the-art algorithms in terms of both solution convergence and diversity. This research provides a novel theoretical perspective and an effective solution for the multi-robot task allocation problem under energy constraints.
\end{abstract}

\begin{IEEEkeywords}
Multi-objective optimization, multi-robot task allocation, payload constraint, energy constraint, task divisibility characteristic

\end{IEEEkeywords}

\section{Introduction}

The fusion of artificial intelligence and robotics is profoundly reshaping numerous traditional sectors and has made significant contributions to enhancing production efficiency and alleviating labor pressures~\cite{chen2024efficient}. In the domain of smart farming, developing efficient automated solutions has become imperative to meet rising crop yields and urgent harvesting demands~\cite{asmara2024economic}. Although various agricultural robots have been deployed to reduce manual intervention~\cite{ranjha2021facilitating}, single-robot systems often prove inefficient when handling large-scale scenarios~\cite{ka2024systematic}. Consequently, multi-robot systems (MRS) that can operate collaboratively have emerged as an inevitable trend in the field.

In agricultural operations, due to individual payload capacities and fleet size limitations, robots typically require multiple trips to complete their assigned tasks. Therefore, designing an MRS involves not only determining the task assignments for each robot but also planning their routes and optimizing the task execution sequence. This class of problem is defined as the agricultural multi-robot task allocation (MRTA) problem, which is NP-hard~\cite{dong2024effective}. Furthermore, addressing conflicting objectives, such as minimizing the overall task completion time (makespan) and transportation cost, adds another layer of complexity to the problem.

Research into the agricultural MRTA problem has followed a clear evolutionary trajectory, with models progressively deepening to better reflect real-world conditions. Pioneering work by Dai et al.~\cite{dai2023multi}, MODABC, first established an algorithm for route construction and task allocation for multiple harvesting robots. Building upon this, subsequent research has focused on improving its search strategies: CDABC~\cite{guo2024effective} enhances performance by specifically targeting the bottleneck robot for weeding work, while AMOEA~\cite{dong2024effective} seeks better team balance by focusing on both the most and least utilized robots for spraying applications. The scope of research has also expanded to the cooperative operation of weeding robots and spraying drones~\cite{wang2024multi}. However, these advanced methods predominantly model each task as an indivisible atomic unit. This modeling choice severely limits the utilization efficiency of robot payload and time in practical scenarios requiring high scheduling flexibility. Although Guo et al.~\cite{guo2024hybrid}  introduced several task-splitting modes to address this limitation, its problem setting (e.g., allowing only a single trip per robot) also restricts its direct applicability in orchard scenarios. To overcome this bottleneck, our prior work~\cite{Chen2025adaptive} significantly improved resource utilization and overall solution quality by introducing a task-splitting mechanism that allows for the flexible distribution of a single task's workload.

However, the aforementioned studies overlook a deeper challenge: energy management. The introduction of various search strategies, while greatly enhancing scheduling flexibility, has also complicated the energy management of robot routes. This has pushed the hard constraint of finite battery capacity to the forefront of the optimization problem, establishing it as a core bottleneck that determines solution quality. The challenge of energy management has been extensively studied in the domain of the electric vehicle routing problem (EVRP). For instance, its energy consumption models have evolved from simplified assumptions of uniform change to more realistic non-linear functions~\cite{bruglieri2023matheuristic}, and its energy replenishment strategies have diversified from the conventional charge-to-full strategy to more flexible options such as partial charging~\cite{bavand2022online} and battery swapping~\cite{tarar2023techno}. Among these, the battery swapping model can save considerable charging and potential queuing time, which can greatly enhance harvesting efficiency during the busy season~\cite{chen2025multiobjectivetaskallocationelectric}. To solve these problems, optimization strategies such as variable neighborhood search~\cite{yilmaz2022variable}, artificial bee colony (ABC)~\cite{guo2023low}, and ant colony optimization (ACO)~\cite{fan2024two} have also been developed. Seeking to transcend the traditional single-objective focus of EVRP research, studies by Zhou~\cite{zhou2022multi} and Comert et al.~\cite{comert2023new} have successfully applied multi-objective optimization (MOO) frameworks to this problem, opening new research avenues for navigating the trade-offs between conflicting objectives like economic cost and temporal efficiency.

Nevertheless, the majority of EVRP studies focus on single-trip logistics and rarely address the unique complexities of agricultural MRTA, such as multi-trip operations with dynamic payloads and task divisibility. Since both unloading and battery swapping for the robots occur at the depot, these operational characteristics create a cascading effect between the energy and payload constraints. Specifically, a mandatory battery swap simultaneously resets the robot's load, which can disrupt the structure of the subsequent task plan and render an optimized solution instantly suboptimal. Consequently, existing methods are difficult to apply directly, leaving the systematic reconstruction of paths disrupted by energy constraints as a critical, unaddressed gap within the agricultural MRTA domain.

To this end, this study focuses on orchard harvesting scenarios and establishes an MOO model that addresses robotic harvesting and transportation tasks by considering the robots' payload constraints, energy constraints, and task divisibility characteristics. Given the difficulty of obtaining optimal solutions with exact algorithms within a reasonable time limit~\cite{choudhury2022dynamic}, this study proposes a heuristic segment anchoring-based balancing algorithm (SABA). Its core component, the sequential anchoring and balancing mechanism (SABM), leverages charging decisions as anchors to systematically reconstruct routes that have been fragmented by energy constraints. Furthermore, the proposed proportional splitting-based rebalancing mechanism (PSRM) assists SABA in performing a fine-grained balancing of the final solutions' makespans. The main contributions of this study can be summarized as follows:
\begin{enumerate}
    \item Formulation of a mathematical model tailored for agricultural MRTA that simultaneously considers task divisibility characteristics and energy constraints;
    \item Development of a novel heuristic algorithm SABA whose `anchor-and-reconstruct' paradigm offers a new approach for handling route reconstruction under energy constraints;
    \item Validation of SABA's comprehensive superiority over 6 state-of-the-art algorithms through experiments on a real-world case study and 15 benchmark instances.
\end{enumerate}
%\vspace{-1em}

\begin{figure}[htp]
    \centering
    \includegraphics[width=8.5cm]{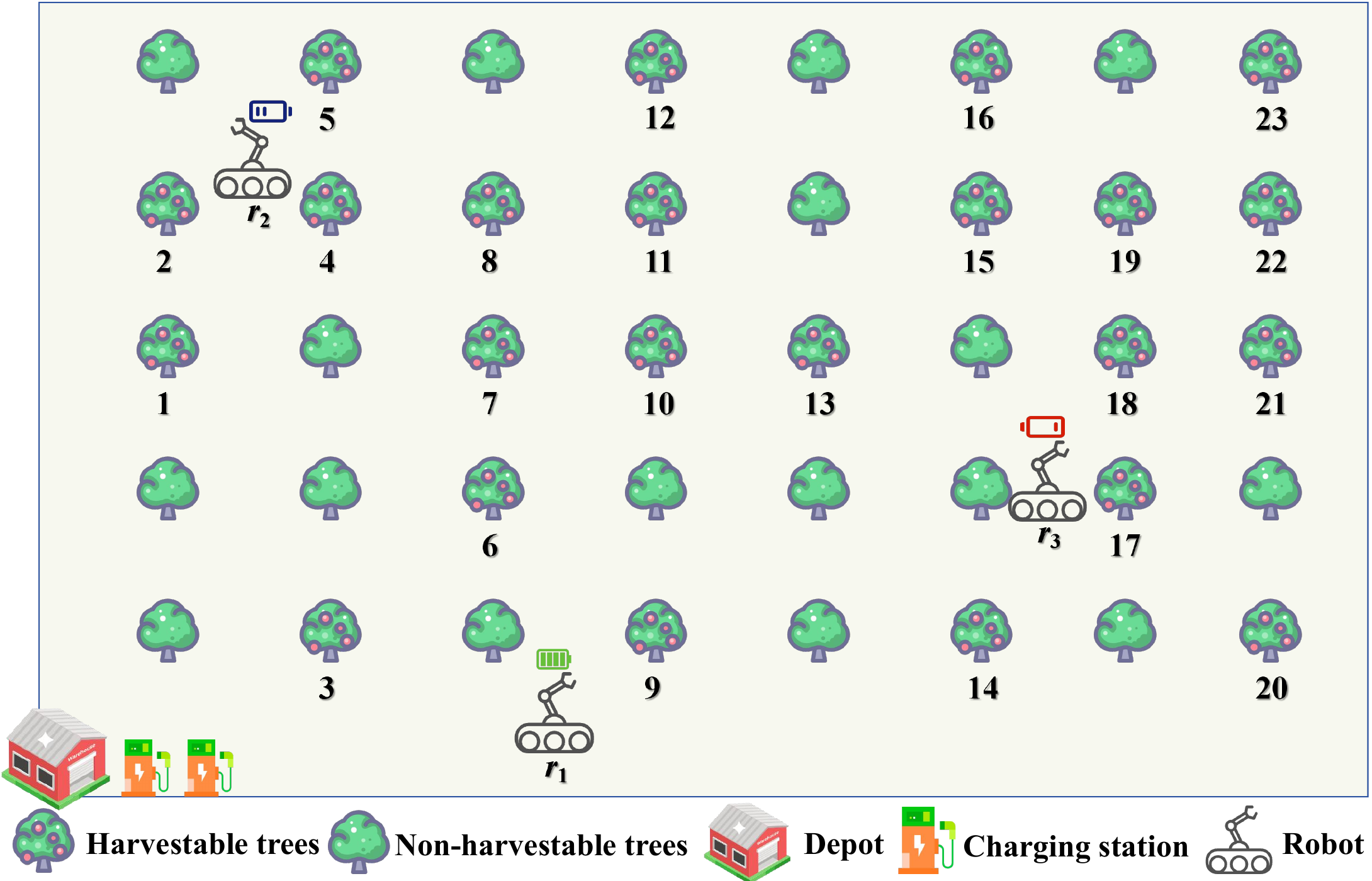}
    \caption{Schematic diagram of the orchard}
    \label{fig_scence}
\end{figure}
%\vspace{-1em}

\section{Proposed Modeling}

\subsection{Problem Description}

This study addresses a complex, multi-robot selective harvesting problem derived from a real-world orchard scenario in Henan Province, China. Within the orchard, even slight variations in environmental factors, such as light exposure and soil fertility, result in non-uniform fruit maturity. Leveraging precision agriculture principles, a robotic team performs batched harvesting operations, targeting only trees with fruit that meets predefined standards for maturity indicators like sugar content and coloration. Figure~\ref{fig_scence} illustrates the spatial layout for a single operational batch, which includes a central depot and $n$ task nodes designated for harvesting. In this setting, the travel distance between any two nodes (with the shortest feasible path) is pre-calculated. The limited payload capacity of the robots necessitates multiple trips to service all assigned tasks. In addition, the finite battery capacity prevents the robots from completing all tasks in an uninterrupted tour. Considering the limited number of spare batteries at the depot, a battery swap is permitted only when a robot's battery level falls below a predefined threshold ($B_{th}$). The primary aim of this research is to simultaneously minimize two conflicting objectives: the makespan to complete all tasks and the total transportation cost of the robotic system.

\subsection{Mathematical Model}
\textbf{Sets and Parameters}
\begin{flalign*}
& N = \{-1,0,1,\ldots,n\}\!\colon \text{ collection of task nodes }&& \\
& R = \{1,\ldots,r\}\!\colon \text{ fleet of available robots}&& \\
& S = \{1,\ldots,s\}\!\colon \text{ set of all possible cycles} && \\
& S^r\!\colon \text{ complete cycle of robot } r && \\
& d_{ij}\!\colon \text{ travel distance from node } i \text{ to } j \text{ (meters, m)} && \\
& q_i\!\colon \text{ yield of fruit at node } i \text{ (number of fruits)} && \\
& Q=300\!\colon \text{ load capacity per robot (number of fruits) } && \\
& W=30\!\colon \text{ weight of an empty robot (kilograms, kg)} &&  \\
& w_a=0.3\!\colon \text{ average weight per fruit (kg)} && \\
& t_{\text{pick}}=7\!\colon \text{ average picking time per fruit (seconds, s)~\cite{xiong2019development}} && \\
& V=1\!\colon \text{ constant travel speed of robots (m/s)~\cite{guo2024hybrid}} && \\
& g=9.81\!\colon \text{ gravitational acceleration (m/s$^2$)~\cite{eckert2012a+}} && \\
& \mu=0.05\!\colon \text{ coefficient of rolling resistance~\cite{valero2017influence}} && \\
& \eta=0.8\!\colon \text{ coefficient of the energy efficiency~\cite{DOE2024EV}} && \\
& B_c=432\!\colon \text{battery capacity of a robot (kilojoules, kJ)~\cite{mcnulty2022review}} && 
\end{flalign*}
\begin{flalign*}
& B_{th}=20\%B_c\!\colon \text{ battery threshold to trigger a swap (kJ)~\cite{mitici2022electric}} && \\
& t_{\text{swap}}=150\!\colon \text{ time required for a battery swap (s)~\cite{yang2015battery}} && \\
& e=0.3\!\colon \text{ average picking energy per fruit (kJ)~\cite{lou2024analysis}} && \\
&  E_{\text{serv},i}\!\colon\text{service energy consumpution on task } i \text{ (kJ)} && \\
& E_{ijk}\!\colon \text{ travel energy for robot } k \text{ from } i \text{ to } j \text{ (kJ)} && \\
& E_s\!\colon\text{ total transportation energy on cycle } s \text{ (kJ)} && \\
& T_{ijk}\!\colon \text{ travel time for robot } k \text{ from node } i \text{ to } j \text{ (s)} && \\
& T_{iks}\!\colon \text{ picking time at node } i \text{ by robot } k \text{ on cycle } s \text{ (s)} && \\
& T_{\text{swap},s}\!\colon \text{ battery swap time incurred on cycle } s \text{ (s)} && \\
& T_s\!\colon \text{ total completion time of cycle } s \text{ (s)} && \\
& T_{\text{max}}\!\colon \text{ maximum completion time (makespan) (s)} &&\\
& E_{\text{total}}\!\colon \text{ total transportation cost} &&
\end{flalign*}

\textbf{Decision Variables}
\begin{flalign}
& x_{ijk} = \begin{cases}
1, & \text{if robot } k \text{ travels from node } i \text{ to node } j \\
0, & \text{otherwise}
\end{cases} && \notag \\
& \qquad \forall i,j \in N, i \neq j, \forall k \in R \nonumber\\[2ex]
& L_{ik} \geq 0: \text{ load of robot } k \text{ when leaving node } i && \notag \\
& \qquad \forall i \in N, \forall k \in R \nonumber\\[2ex]
& B_{ik} \geq 0: \text{ battery level of robot } k \text{ when reaching node } i && \notag \\ 
& \qquad \forall i \in N, \forall k \in R \nonumber\\[2ex]
& p_{iks} \in [0,1]: \text{ completion proportion of task } i \text{ on cycle } s && \notag \\
& \qquad \forall i \in N \setminus \{-1,0,1\}, \forall k \in R, \forall s \in S \nonumber\\[2ex]
& z_{ks} = \begin{cases}
1, & \text{if robot } k \text{ executes cycle } s \\
0, & \text{otherwise}
\end{cases} && \notag \\
& \qquad \forall k \in R, \forall s \in S \nonumber
\end{flalign}

\textbf{Energy and Time Components}
\begin{flalign}
& T_{ijk} = \frac{d_{ij}}{V} \qquad \forall i,j \in N, i \neq j, \forall k \in R&& \notag \\[2ex]
& T_{iks} = \begin{cases}
 t_{\text{pick}} \cdot q_i \cdot p_{iks}, & i \in N \setminus \{-1,0,1\} \\
0, & \text{otherwise}
\end{cases} && \notag \\
& \qquad \forall i \in N, \forall k \in R, \forall s \in S \nonumber \\[2ex]
& T_{\text{swap},s} = \begin{cases} t_{\text{swap}}, & \text{if a battery swap occurs on cycle } \\ 0, & \text{otherwise} \end{cases} && \notag \\
& \qquad \forall s \in S \nonumber\\[0.5ex]
& T_s = \sum_{k \in R}z_{ks}\left(\sum_{i \in N}\sum_{j \in N, j \neq i} T_{ijk} \cdot x_{ijk} + \sum_{i \in N}T_{iks} \right) + T_{\text{swap},s} && \notag \\[2ex]
& E_{ijk} = \frac{d_{ij}(W + L_{ik} \cdot w_a)g\mu}{\eta} \times 10^{-3} && \notag \\
& \qquad \forall i,j \in N, i \neq j, \forall k \in R \nonumber\\[2ex]
& E_{\text{serv}, iks} = (q_i \cdot p_{iks}) \cdot e && \notag \\
& \qquad \forall i \in N \setminus \{-1,0,1\}, \forall k \in R, \forall s \in S \nonumber
\end{flalign}
\begin{flalign}
& E_s = \sum_{k \in R}z_{ks}\left(\sum_{i \in N}\sum_{j \in N, j \neq i} E_{ijk} \cdot x_{ijk}\right) && \notag
\end{flalign}

\textbf{Objective Functions}
\begin{flalign}
&\begin{aligned}
& \min \text{ } T_{\text{max}} = \max_{k \in R} \sum_{s \in S} T_s \cdot z_{ks} \\
& \parbox[t]{1\columnwidth}{\small Minimizes the system makespan.}
\end{aligned} && \notag \\[3ex]
&\begin{aligned}
& \min \text{ } E_{\text{total}} = \sum_{s \in S}E_s \\
& \parbox[t]{1\columnwidth}{\small Minimizes the total transportation cost, represented by the energy consumed during travel.}
\end{aligned} && \notag
\end{flalign}

\textbf{Constraints}
\begin{flalign}
&\begin{aligned}
& \sum_{k \in R}\sum_{s \in S} p_{iks} = 1 \qquad \forall i \in N \setminus \{-1,0,1\} \\
& \parbox[t]{1\columnwidth}{\small Ensures that the demand of each task is fully met, allowing for workloads to be split.}
\end{aligned} && \notag \\[3ex]
&\begin{aligned}
& q_i \cdot p_{iks} \in \mathbb{Z}_{\ge 0} \qquad \forall i \in N \setminus \{-1,0,1\}, \forall k \in R, \forall s \in S \\
& \parbox[t]{1\columnwidth}{\small Guarantees that the number of fruits picked for any given proportion is an integer.}
\end{aligned} && \notag \\[3ex]
&\begin{aligned}
& L_{jk} = 0 \qquad \forall j \in \{-1,0,1\}, \forall k \in R \\
& \parbox[t]{1\columnwidth}{\small Resets a robot's load to zero upon each visit to the depot.}
\end{aligned} && \notag \\[3ex]
&\begin{aligned}
& L_{jk} = \sum_{i \in N \setminus \{j\}}\sum_{s \in S} (L_{ik} + q_j \cdot p_{jks}) \cdot x_{ijk} \cdot z_{ks}  \\
& \qquad \forall j \in N \setminus \{-1,0,1\}, \forall k \in R \\
& \parbox[t]{1\columnwidth}{\small Defines the load propagation, updating a robot's load after it services a task.}
\end{aligned} && \notag \\[3ex]
&\begin{aligned}
& L_{ik} \leq Q \qquad \forall i \in N, \forall k \in R \\
& \parbox[t]{1\columnwidth}{\small Enforces the maximum load capacity for each robot.}
\end{aligned} && \notag \\[3ex]
&\begin{aligned}
& B_{jk} = B_{ik} - E_{ijk} - E_{\text{serv}, iks} \\
& \qquad \forall i,j \in N, i \neq j, \forall k \in R, \forall s \in S \\
& \parbox[t]{1\columnwidth}{\small Models the battery dynamics, where the battery level decreases due to both travel and service energy.}
\end{aligned} && \notag \\[3ex]
&\begin{aligned}
& 0 \leq B_{ik} \leq B_c \qquad \forall i \in N, \forall k \in R \\
& \parbox[t]{1\columnwidth}{\small Ensures the battery level remains within its physical limits (zero to full capacity).}
\end{aligned} && \notag \\[3ex]
&\begin{aligned}
& \sum_{s \in S}  T_s \cdot z_{ks} \leq T_{\text{max}} \qquad \forall k \in R \\
& \parbox[t]{1\columnwidth}{\small Links the completion time of each robot to the overall makespan objective, $T_{\text{max}}$.}
\end{aligned} && \notag \\[3ex]
&\begin{aligned}
& \sum_{k \in R}  z_{ks} \leq 1 \qquad \forall s \in S \\
& \parbox[t]{1\columnwidth}{\small Ensures that each planned route is assigned to at most one robot.}
\end{aligned} && \notag
\end{flalign}
where:
\begin{itemize}
    \item A battery swap is performed if and only if a robot returns to the depot (node $i \in \{-1,0,1\}$) with a battery level $B_{ik} \leq B_{th}$.
    \item To ensure route validity, the formulation incorporates Miller-Tucker-Zemlin MTZ constraints to prevent the formation of subtours disconnected from the depot~\cite{miller1960integer}. For the sake of brevity, other standard auxiliary constraints are not explicitly detailed here.
    \item For the purpose of conducting realistic numerical simulations, all parameter values within the model have been calibrated based on established findings in the literature and practical operational data~\cite{dai2023multi}. In addition, the parameters are illustrative and may vary in different applications.
\end{itemize}

\begin{figure*}[htp]
    \centering
    \includegraphics[width=18.2cm]{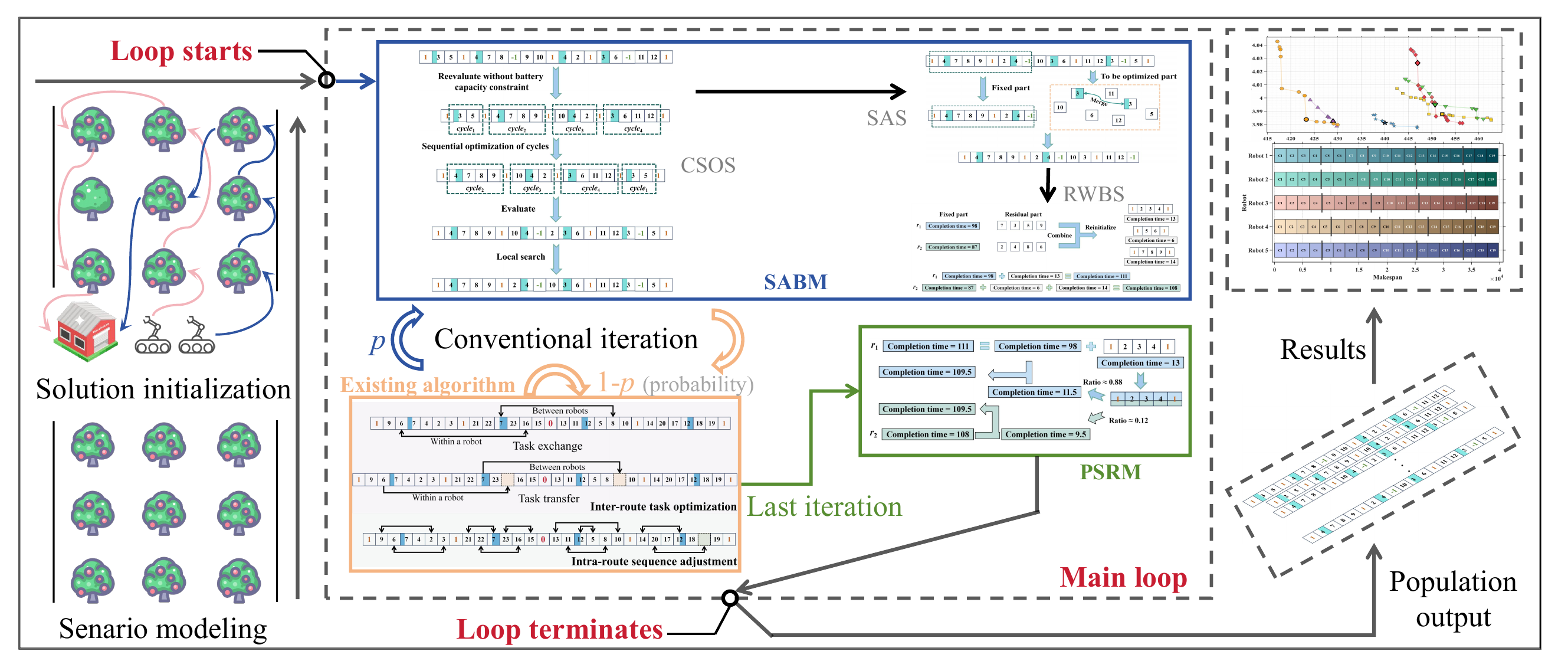}
    \caption{Main procedure of SABA}
    \label{fig:main}
\end{figure*}

%\vspace{-0.5em}
\section{Methodology}

The proposed SABA is a significant extension of our prior work~\cite{Chen2025adaptive}. It is specifically designed to address the complex cascading effect caused by energy constraints. As illustrated in Fig.~\ref{fig:main}, SABA achieves this by incorporating two novel core mechanisms into the existing algorithm: the sequential anchoring and balancing mechanism (SABM) and the proportional splitting-based rebalancing mechanism (PSRM). 

The algorithm first employs SABM to repair the initial solutions, whose quality degrades due to the introduction of energy constraints. It performs a systematic reconstruction and optimization on the initial plans that have become structurally suboptimal due to forced recharging, aiming to minimize the negative impacts of charging decisions on the overall efficiency of the schedule. Subsequently, during the main evolutionary loop of SABA, the population primarily evolves using the existing operators~\cite{Chen2025adaptive}. Given that SABM is computationally intensive, it functions as a structural optimization operator, invoked with a probability $p$, to guide the population towards more promising, structured regions of the search space without imposing a significant computational burden. To make full use of the computational budget, SABA dynamically monitors its iteration time. When the estimated remaining time is insufficient to complete another full generation, PSRM is initiated. This final mechanism performs a fine-grained adjustment and balancing of the robot makespans within the current set of elite solutions. Ultimately, SABA outputs a set of non-dominated solutions to provide decision-makers with a diverse range of high-quality scheduling plans. The overall process of the algorithm is shown in Algorithm~\ref{alg:saba_framework}. Due to space limitations, the computational complexity and pseudocode for other mechanisms are presented in the Appendix.

\begin{algorithm}[htb]
\SetAlgoLined
\KwIn{%
    \begin{tabular}[t]{ll}
        Task set: & $N$\\
        Number of robots: & $r$\\
        Population size: & $P_{\text{size}}$\\
        Time limit: & $T_{\text{max}}$\\
        SABM invocation probability: & $p$\\
        Problem parameters: & $params$
    \end{tabular}
}
\vspace{0.1em}
\KwOut{Final population, $P_{\mathrm{final}}$}
\vspace{0.1em}
$current\_time \leftarrow 0$\;
$iter \leftarrow 0$\;
\vspace{0.1em}
$P \leftarrow \text{Solution initialization}$\;
\vspace{0.1em}
\tcp{Main Evolutionary Loop}
\While{$current\_time \le T_{\textnormal{max}}$}{
\tcp{Initial repair with SABM}
\If{$iter = 0$}{
Update $P$ with SABM\;}
    Update $P$ with existing algorithm \tcp{Population evolution}
    \vspace{0.1em}
    \If{$\text{rand}() \le p$}{
        Update non-dominated solutions in $P$ with SABM\;
    }
    \vspace{0.1em}
    \tcp{Check for final refinement with PSRM}
    $iter \leftarrow iter + 1$\;
    Get current time $current\_time$\;
    $iter\_time \leftarrow current\_time / iter$\;
    \If{$current\_time + iter\_time \ge T_{\textnormal{max}}$}{
        Update non-dominated solutions in $P$ with PSRM\;
        \textbf{break} \tcp{Terminate loop for final output}
    }
}
\vspace{0.1em}
$P_{\mathrm{final}} \leftarrow P$ \tcp{Output final popualtion}
\Return{$P_{\mathrm{final}}$}
\caption{The SABA Framework}
\label{alg:saba_framework}
\end{algorithm}

\begin{figure}[htp]
    \centering
    \includegraphics[width=9cm]{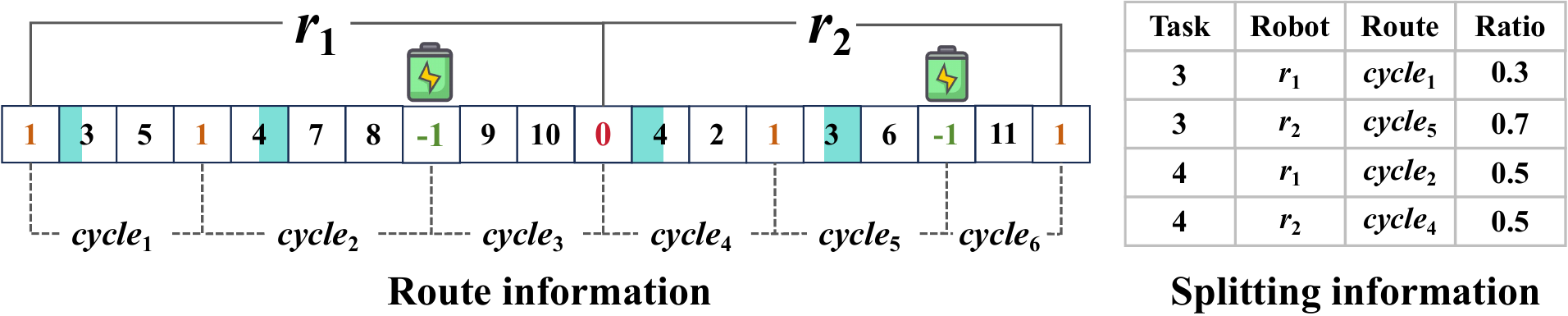}
    \caption{Solution representation diagram}
    \label{fig:solution}
\end{figure}
% %\vspace{-2.5em}
%\vspace{-2.5em}
\subsection{Solution representation}

To accommodate the split-delivery nature of the tasks, SABA employs a hybrid encoding scheme comprising route information and splitting information, as illustrated in Fig.~\ref{fig:solution}.

The route information is represented as an integer vector that defines the global execution sequence of all tasks. In addition to task nodes, this vector includes three special delimiters with distinct semantics to represent visits to the depot. The `0' marker serves as a robot delimiter, demarcating the task sequences assigned to different robots. `1' indicates a load-induced depot visit, triggered when the vehicle reaches its maximum load capacity. Finally, `-1' represents an energy-induced depot visit, necessitated by insufficient energy; this visit not only resets the load but also replenishes the battery, incurring an additional battery swap time.

The splitting information ensures that since all tasks within a single cycle are distinct, the cycle itself can serve as a label to differentiate each split task. This allows for the direct indexing of each task's corresponding service ratio during the solution evaluation. This design decouples the topological structure of the routes from the service quantity of each split task. During the evolutionary process, whenever genetic operators modify the route vector, the splitting information must be adjusted synchronously to maintain solution integrity.

Compared to a direct solution representation method~\cite{dai2023multi}, the current approach can generate a clearer mapping of robot paths and exhibits lower complexity than hierarchical hybrid encoding schemes~\cite{10964773}.

%\vspace{-1.7em}

\subsection{Sequential anchoring and balancing mechanism}
\label{S-SABM}

\begin{figure*}[htp]
    \centering
    \includegraphics[width=18cm]{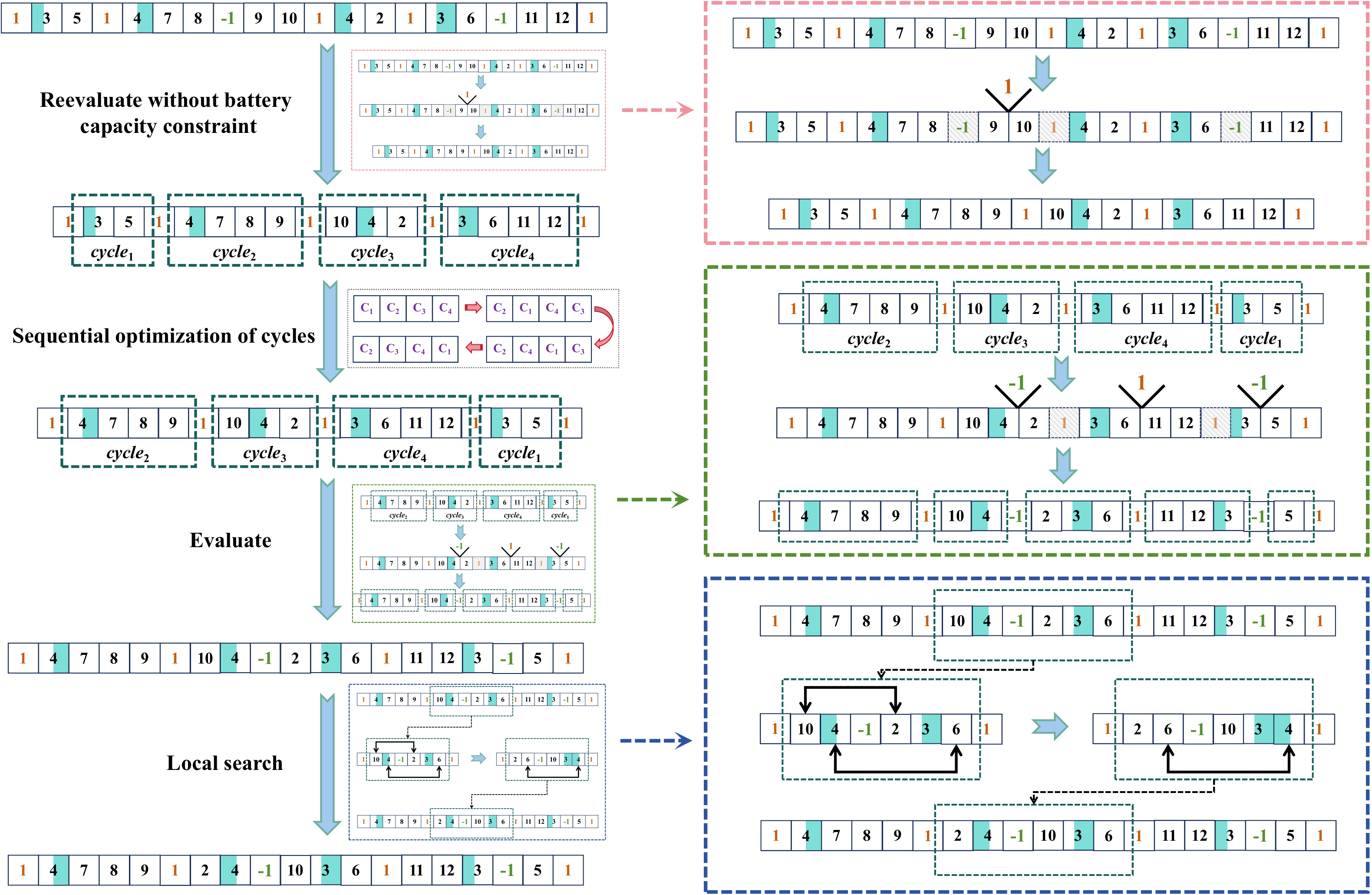}
    \caption{Schematic diagram of CSOS}
    \label{fig_CSOS}
\end{figure*}

SABM is a core repair and enhancement mechanism, designed to systematically reconstruct solutions whose structures have been degraded by the introduction of energy constraints. During the evolutionary process, SABM is selectively applied only to the non-dominated set of the population. This elite strategy aims to efficiently improve the quality of the best individuals, thereby guiding the evolution of the entire population. SABM consists of three key steps:

\subsubsection{Cycle sequence optimization step}

The cycle sequence optimization step (CSOS) aims to mitigate the negative impacts of forced recharging on the overall plan through macroscopic adjustments. As the algorithm's initialization and standard evolutionary operators primarily operate at the cycle level, the execution order of these cycles is often suboptimal after energy constraints are considered. Therefore, as shown in Fig.~\ref{fig_CSOS}, CSOS first decomposes each robot's full path into a set of independent operational cycles.

Subsequently, the algorithm treats this as a permutation optimization subproblem. We employ the classic 2-opt algorithm~\cite{englert2014worst} to reorder the execution sequence of these cycles to find a superior macroscopic sequence. This allows CSOS to strategically position geographically distant or task-heavy cycles at more favorable places, creating better preconditions for subsequent charging decisions at a global level and reducing unnecessary long-distance returns to the depot.

Finally, the task sequence within the specific cycle segment disrupted by a charging insertion is adjusted by local search. More complex tasks within the current cycle are strategically placed before the charging point to form a separate cycle, while the remaining tasks are subject to further optimization, thus reducing the burden on subsequent operations.

\subsubsection{Segment anchoring step}

\begin{figure}[htp]
    \centering
    \includegraphics[width=9cm]{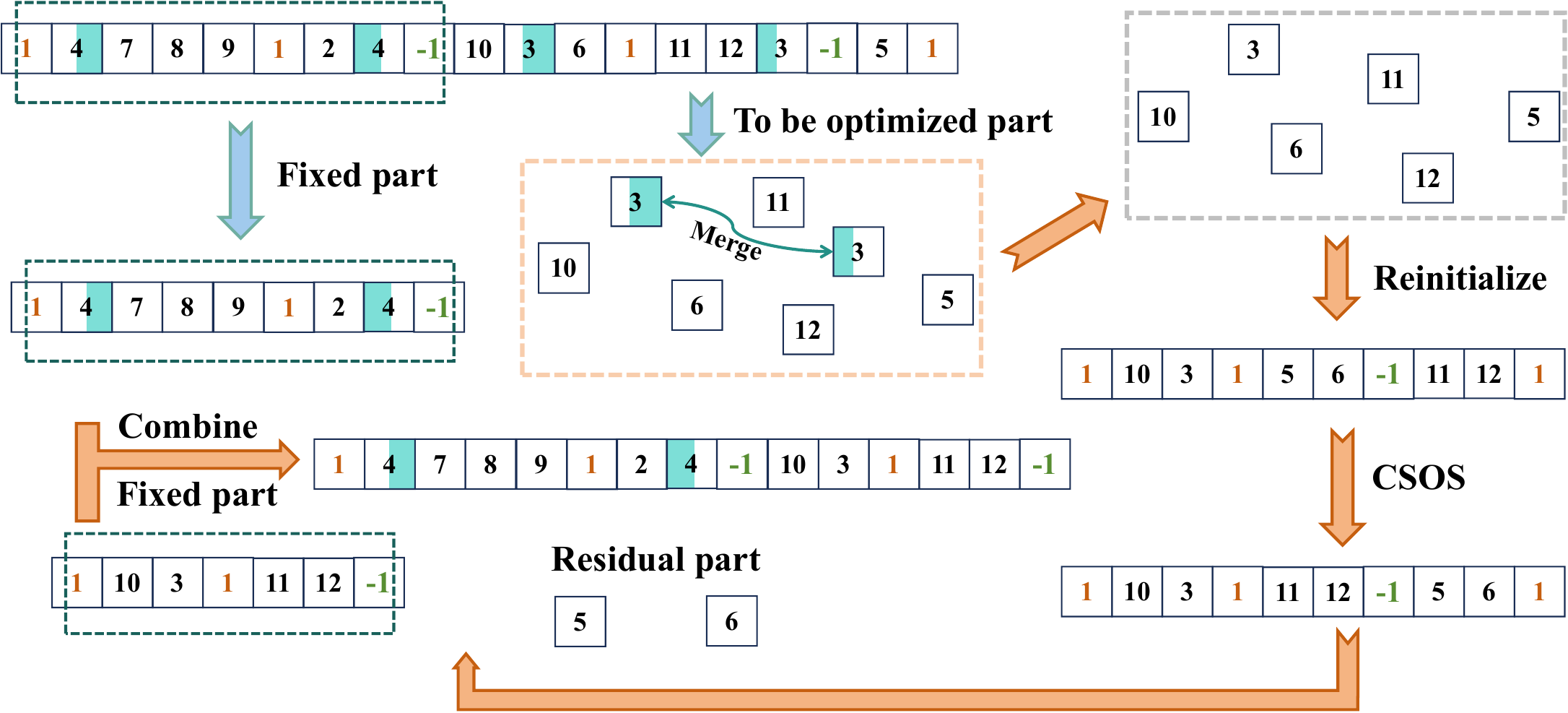}
    \caption{Schematic diagram of SAS}
    \label{fig_SAS}
\end{figure}

After CSOS determines the macroscopic cycle sequence, the segment anchoring step (SAS) performs an iterative, coarse-to-fine micro-structural optimization. The core idea is to use the energy-induced charging stops (-1) as anchor points to progressively solidify and refine the path structure. As illustrated in Fig.~\ref{fig_SAS}, SAS first identifies the first occurrence of a charge stop (-1) in the current route. The portion of the route from the start to this point is designated as an anchored segment and is locked from further changes in subsequent SAS iterations. Then, all residual tasks after the anchored segment are treated as a new subproblem to be planned. The algorithm reinitializes this small-scale subproblem scenario to generate a new, optimized sequence of tasks for them. Subsequently, CSOS is invoked again on the newly generated sequence, and another anchored fragment is created. This process continues cyclically until the remaining tasks are insufficient to trigger a new charging stop. Finally, all generated anchored fragments are stitched together in their order of creation. Through this cyclical anchoring and reconstruction, the SAS transforms a path, initially fragmented by charging needs, into a structurally sound and logically coherent plan.

\subsubsection{Residual workload balancing step}

After the SAS, each robot's path structure is relatively stable, but significant makespan imbalances may still exist across the team. The objective of the residual workload balancing step (RWBS) is to address this issue through global resource redistribution.

\begin{figure}[htp]
    \centering
    \includegraphics[width=9cm]{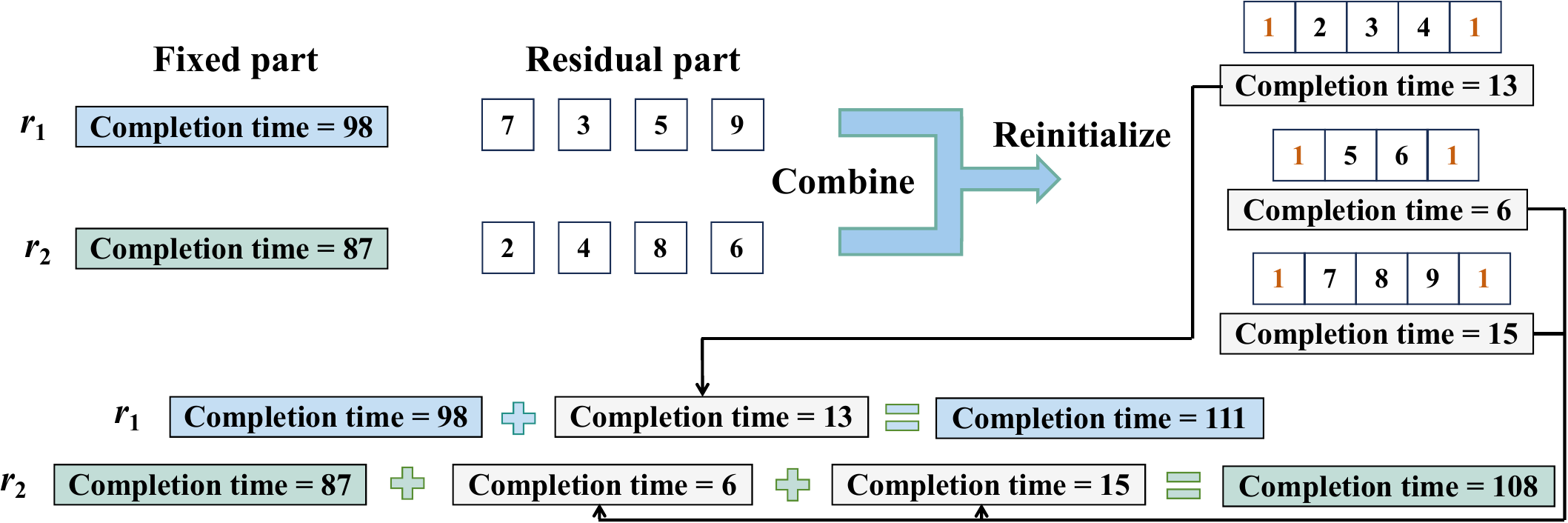}
    \caption{Schematic diagram of RWBS}
    \label{fig_RWBS}
\end{figure}

RWBS first identifies the residual tasks for each robot, defined as those appearing after its final charging anchor point. These tasks are chosen for reallocation because their position at the end of the schedule minimizes the impact on the already anchored path structure. The residual tasks from all robots, along with their corresponding remaining demands, are collected into a common task pool, where all instances of the same task are merged. The algorithm then treats this task pool as an another independent, smaller-scale subproblem and re-plans it into a new set of efficient operational cycles. These newly generated cycles are then assigned to the different robots based on their estimated durations, aiming to close the work-time gap based on their existing, anchored workloads, and thereby minimizing the final makespan. A schematic diagram of this process is shown in Figure~\ref{fig_RWBS}.

\subsection{Proportional splitting-based rebalancing mechanism}

As task splitting can significantly enlarge the search space and hinder the speed of population evolution~\cite{Chen2025adaptive}, PSRM is applied in the final stage of the SABA, serving as a fine-tuning step. This mechanism is applied to each solution in the final Pareto-optimal set at the last iteration. It aims to further reduce the team's maximum completion time ($T_{\rm{max}}$) through fine-grained task splitting and reallocation, thereby enhancing the temporal efficiency and balance of the schedules without significantly increasing the total transportation cost.

For a given schedule, PSRM identifies the bottleneck robot ($r_b$) with the longest makespan, which defines the current $T_{\rm{max}}$. It then selects a suitable donor cycle from this robot's path. The selection follows two principles: 1) it must be a cycle executed after $r_b$'s final charge to protect the main path structure solidified by SABM; and 2) the cycle with the lowest transportation cost is chosen to minimize the extra energy overhead from duplicated travel.

\begin{figure}[htp]
    \centering
    \includegraphics[width=9cm]{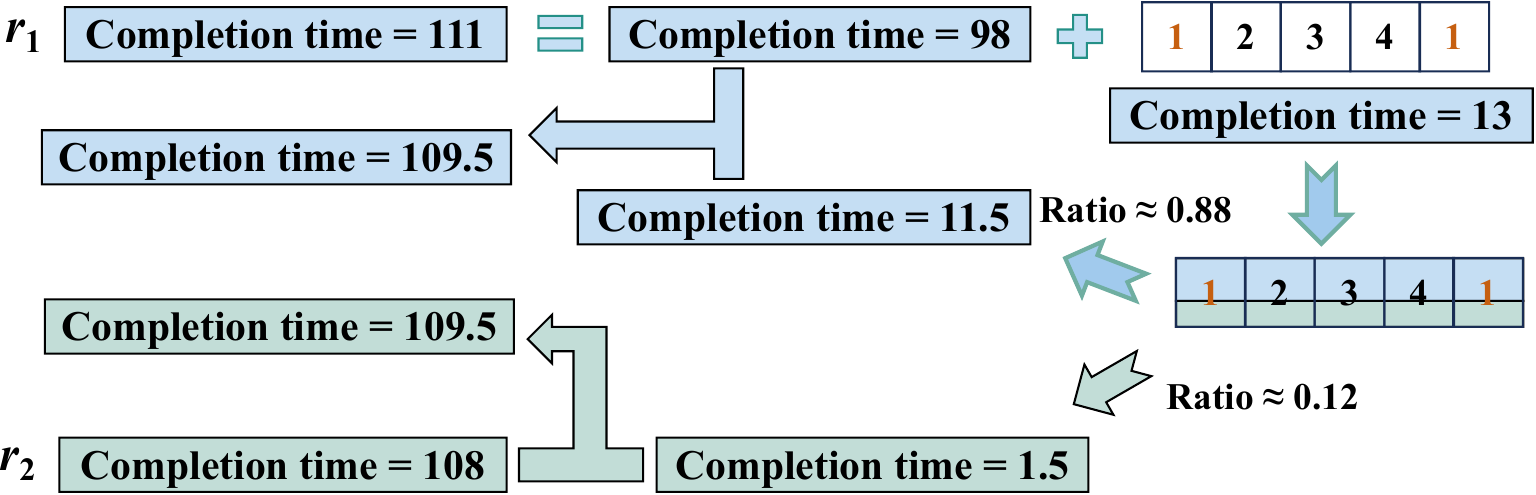}
    \caption{Schematic diagram of PSRM}
    \label{fig_PSRM}
\end{figure}

As shown in Fig.~\ref{fig_PSRM}, PSRM then precisely splits the donor cycle based on the makespan gaps between the different robots. The resulting new cycles differ from the donor cycle only in their task completion ratios. By repeating this process for multiple robots in the team, PSRM smoothly transfers the workload from the most burdened robot to underutilized ones in a fine-grained, controlled manner. Such granular adjustments are difficult to achieve with traditional operators that move entire cycles, and ultimately, this mechanism can significantly improve the balance of the solution set, providing the decision-maker with more time-efficient scheduling options.

\section{Experimental results and analysis}

To evaluate the performance of SABA, we conducted a comparative study against six state-of-the-art algorithms: MODABC~\cite{dai2023multi}, AMOEA~\cite{dong2024effective}, RNSGA~\cite{10964773}, AMTSA~\cite{Chen2025adaptive}, HRRA~\cite{chen2025multiobjectivetaskallocationelectric}, and an adapted version of HACO~\cite{comert2023new}. This set of peer algorithms was curated to form a multi-faceted, high-standard benchmark. As pioneering works in agricultural MRTA, MODABC and AMOEA have achieved broad success, establishing them as foundational contributions to the field. Seeking to better reflect real-world applications, subsequent research has incorporated more complex characteristics, with our foundational work, AMTSA, introducing task splitting, and HRRA making preliminary explorations into optimizing charging decisions. Furthermore, we included RNSGA, an emerging and effective algorithm for MRTA problems, to broaden the scope of our comparative analysis. Finally, HACO is a pioneering work from the field of multi-objective EVRP. As the original HACO employs a weighted-sum approach, we developed an enhanced version, HACO$+$, for a fair Pareto-based comparison. This adaptation preserves HACO's core concepts while integrating the environmental selection mechanism from AMTSA and incorporating tailored improvements to its local search operators to better suit our problem context.

All experiments were conducted on a personal computer with a 2.1~GHz Intel Core i7-12700 processor and 32~GB of RAM. To ensure a fair comparison, a uniform termination criterion was applied to all algorithms, defined by a maximum CPU runtime of $T_{\text{max}} = n \times 0.5$ seconds, where $n$ is the number of tasks. The population size for all compared algorithms was set to $P_{\text{size}} = 30$~\cite{dai2023multi}.

%\vspace{-1em}
\subsection{Solution evaluation}

The goal of MOO is to obtain a set of non-dominated solutions, denoted as $\textit{PS}$, which represents the trade-offs among conflicting objectives. Consequently, evaluating an algorithm's performance can hardly rely on individual objective values alone, necessitating a comprehensive metric to assess the overall quality of the solution set~\cite{chen2024archive}. To this end, we utilize the widely-adopted Hypervolume (HV) indicator~\cite{10964773} to measure the quality of the obtained solution sets, with a reference point set to $(1,1)$.

Based on the Lebesgue measure $\delta$, the HV indicator quantifies the volume of the objective space dominated by the entire solution set $\textit{PS}$. It is defined as the union of the hypercubes formed by each solution in $\textit{PS}$ and the reference point:
\begin{equation}
\text{HV}(\textit{PS}) = \delta\left(\bigcup_{j=1}^{|\textit{PS}|} v_j\right)
\end{equation}
where $v_j$ is the hypercube of the $j$-th solution in $\textit{PS}$. A higher HV value generally signifies superior overall performance, as it indicates that the solution set has achieved a better combination of both convergence (closeness to the true Pareto front (PF)) and diversity (distribution of solutions).

In practice, a decision-maker can select a final solution from $\textit{PS}$ based on their specific preference. However, providing a default solution is highly valuable when such preferences are unavailable or undefined. The knee point~\cite{Kaplan2025Knee}, which represents a balanced compromise, resides in the region of the PF with the sharpest trade-off, where a marginal improvement in one objective would cause a significant sacrifice in another. Accordingly, our subsequent comparative experiments will evaluate algorithm performance from two perspectives: 1) non-dominated solution sets, to assess the overall optimization capability; and 2) default output solutions, to assess the ability to provide a high-quality compromise solution.

\subsection{Experimental instances}
\label{s:problem}
To make the evaluation comprehensive, our experimental design includes both a real-world case study and an extended benchmark suite.

First, to validate the practical performance of the algorithms, we utilize a case study based on data from a real-world orchard in Henan Province, China. The orchard comprises approximately 880~apple trees, with an inter-row spacing of 3~m and an intra-row spacing of 2~m. The fruit yield per tree ranges from 30 to 50 apples, with each apple having an average weight of 0.3~kg. The depot, which serves as the central point for both unloading and battery swapping, is positioned 5~m in front of the first row of trees. In the context of a single harvesting batch, we assume that 75\% of the trees meet the harvesting criteria and that a homogeneous fleet of 5~fully charged robots is available for dispatch.

Second, to further assess the generality and robustness of the algorithms across various scales and complexities, we employ a benchmark suite~\cite{Chen2025adaptive}, which consists of 15~test instances generated based on real-world orchard data. As detailed in Table~\ref{tab:instances}, the key parameters for each instance were randomly generated, including the orchard dimensions ($\textit{size}$), the number of tasks ($n$), and the number of available robots ($r \in \{4,5,6,7\}$). $O$ denotes the total fruit yield for a instance, and $D$ represents the average distance from all task locations to the depot.

\begin{table}[]
\centering
\caption{Introduction of problem scenarios}
\setlength{\tabcolsep}{12pt}
\begin{tabular}{cccccc}
\hline
Problem & $\textit{size}$    & $n$    & $r$ & $O$     & $D$  \\
\hline
Pro1    & 30$\times$50   & 140  & 4 & 5645  & 21 \\
Pro2    & 40$\times$70   & 120  & 5 & 4885  & 30 \\
Pro3    & 40$\times$90   & 180  & 4 & 7275  & 48 \\
Pro4    & 60$\times$100  & 560  & 4 & 22150 & 42 \\
Pro5    & 50$\times$120  & 600  & 6 & 24151 & 67 \\
Pro6    & 60$\times$130  & 360  & 6 & 14496 & 49 \\
Pro7    & 60$\times$140  & 420  & 5 & 16706 & 77 \\
Pro8    & 70$\times$150  & 640  & 7 & 25540 & 75 \\
Pro9    & 70$\times$160  & 1260 & 5 & 50155 & 81 \\
Pro10   & 80$\times$170  & 1600 & 5 & 64195 & 82 \\
Pro11   & 90$\times$170  & 1320 & 7 & 52798 & 64 \\
Pro12   & 80$\times$180  & 1680 & 6 & 67568 & 98 \\
Pro13   & 90$\times$190  & 1820 & 7 & 72527 & 70 \\
Pro14   & 100$\times$190 & 1120 & 5 & 44675 & 74 \\
Pro15   & 100$\times$200 & 1500 & 5 & 60197 & 76 \\
\hline
\end{tabular}
\label{tab:instances}
\end{table}

\subsection{Parameter sensitivity and component effectiveness analysis}

Due to page limitation, this part is presented in the Appendix.

\subsection{Results and analysis on a real-world case study}
\label{s:realworld}

\begin{figure}[htp]
    \centering
    \includegraphics[width=8.7cm]{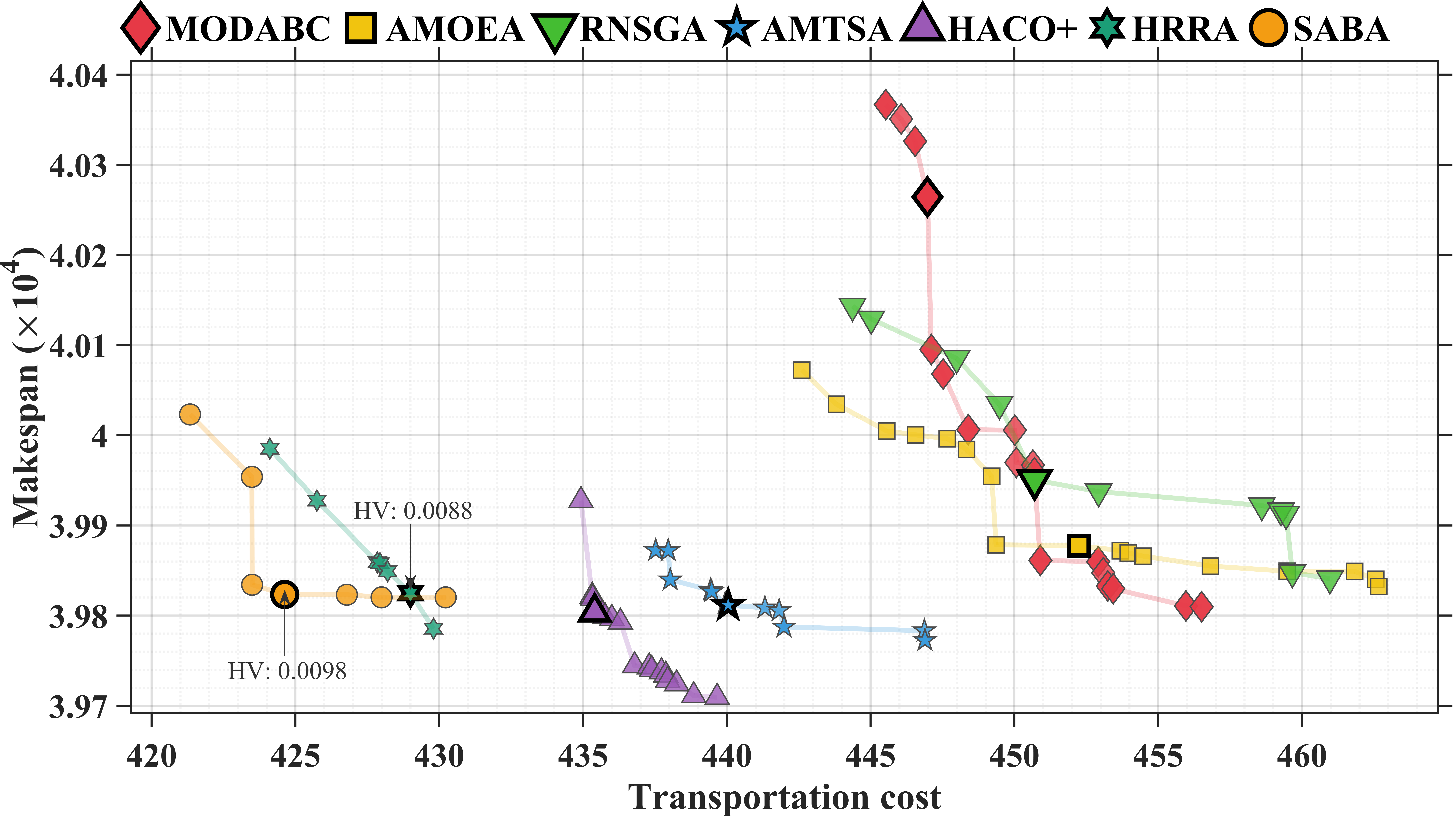}
    \caption{Algorithm performance comparison on the realistic orchard instance}
    \label{fig:PFreal}
\end{figure}

To intuitively compare the algorithms' performance on the real-world instance, Fig.~\ref{fig:PFreal} illustrates the approximate PFs obtained from a typical run. The plot clearly reveals the trade-off achieved by each algorithm between the conflicting objectives of makespan and transportation cost.

A prominent phenomenon is that the solutions from all algorithms naturally form three distinct performance tiers. The third-tier algorithms, MODABC, AMOEA, and RNSGA, converge to a high-cost region due to the lack of specialized designs for the complex constraints. In contrast, the second-tier algorithms demonstrate significant improvement: AMTSA enhances the flexibility of workload distribution via its fine-grained task-splitting mechanism, while the two-stage HACO$^+$ leverages its energy-aware initial search to obtain high-quality starting solutions. Finally, the top tier, consisting of SABA and HRRA, achieves a significantly lower-cost front by incorporating structural optimization for energy constraints and charging decisions. Among them, SABA (represented by yellow circles) exhibits the strongest performance. Its obtained PF largely dominates the solutions from all other algorithms, meaning SABA can achieve a shorter makespan for the same cost, or a lower cost for the same makespan. The points highlighted with a black border represent the default compromise solution from each algorithm, where SABA again holds a leading position. To quantify this comprehensive superiority, the HV indicator of the default output solution generated by SABA is approximately 11\% higher than that of its strongest competitor, HRRA. This overall performance validates its ability to provide decision-makers with a richer portfolio of superior scheduling options for this class of complex problems.

\begin{figure}[h]
    \centering
    \includegraphics[width=8.7cm]{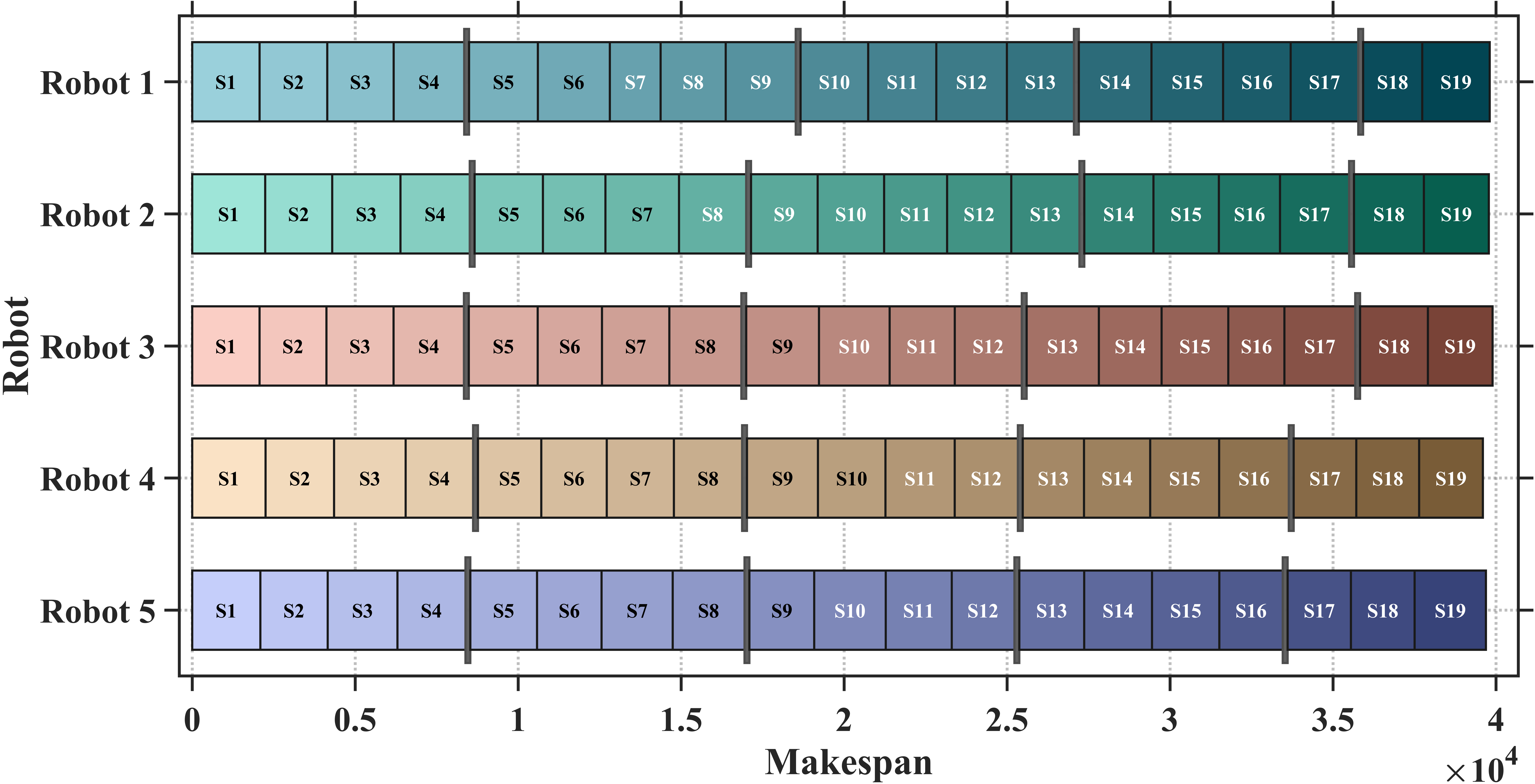}
    \caption{Gantt chart visualization for the default SABA solution}
    \label{fig:Gant}
\end{figure}

To provide a more granular view of the operational schedule generated by SABA, a Gantt chart for its default solution is presented in Fig.~\ref{fig:Gant}. In the chart, each colored horizontal bar represents an individual robot, and the blocks within each bar, labeled $S_i$, denote the $i$-th operational cycle executed by that robot, with the block's length corresponding to its duration. The black delimiter denotes a charging stop. This Gantt chart visually confirms SABA's effectiveness in workload balancing, with the closely aligned completion times of all robots maximizing concurrency and preventing bottlenecks to significantly reduce the team's overall makespan. Additionally, further optimization by the PSRM could achieve an even tighter balance of completion times, but this would come at the cost of increased travel distance from splitting a single cycle into multiple trips. Such a result corresponds to the non-dominated solutions located to the lower-right of the default compromise solution in the Fig.~\ref{fig:PFreal}, representing a trade-off for a lower makespan at a higher cost.

The significant performance disparities primarily reflect the varying capabilities of the algorithms to tackle the problem's core challenges. The algorithms MODABC, AMOEA, and RNSGA, while utilizing different search strategies like adaptive operators or path-first assignment, all lack specialized optimizations for task splitting and charging decisions. Consequently, these algorithms converge to a high-cost region of the objective space, as the simplistic schedules they generate are severely compromised by the unaddressed energy constraints. Although AMTSA achieves a significant performance leap through its flexible task-splitting mechanism, its potential is ultimately limited by a structural neglect of energy constraints, as it lacks any proactive mechanism to manage the disruptions caused by recharging. HACO$^+$, through the global exploration capability of its first stage (ACO), has the opportunity to converge very quickly to a high-quality solution region, thus gaining a significant head-start advantage. However, its two stages lack tight integration, and its second stage also lacks in-depth optimization for charge insertions, leading to high performance volatility. HRRA achieves competitive results by intelligently reallocating tasks after each robot's final charge, which mitigates makespan imbalance to some extent. This optimization, however, is local and end-of-pipe. So it fails to resolve the cascading disruptions caused by earlier charging events in the schedule. In contrast, SABA, through its CSOS, SAS, and RWBS, performs a thorough, multi-level structural reconstruction of paths fragmented by energy constraints. This is followed by the PSRM, which applies a fine-grained makespan balancing. It is this comprehensive design, which deeply couples task splitting with energy optimization, that constitutes the fundamental reason for SABA's comprehensive dominance over all peer algorithms.

\begin{figure}[htp]
    \centering
    \includegraphics[width=8.7cm]{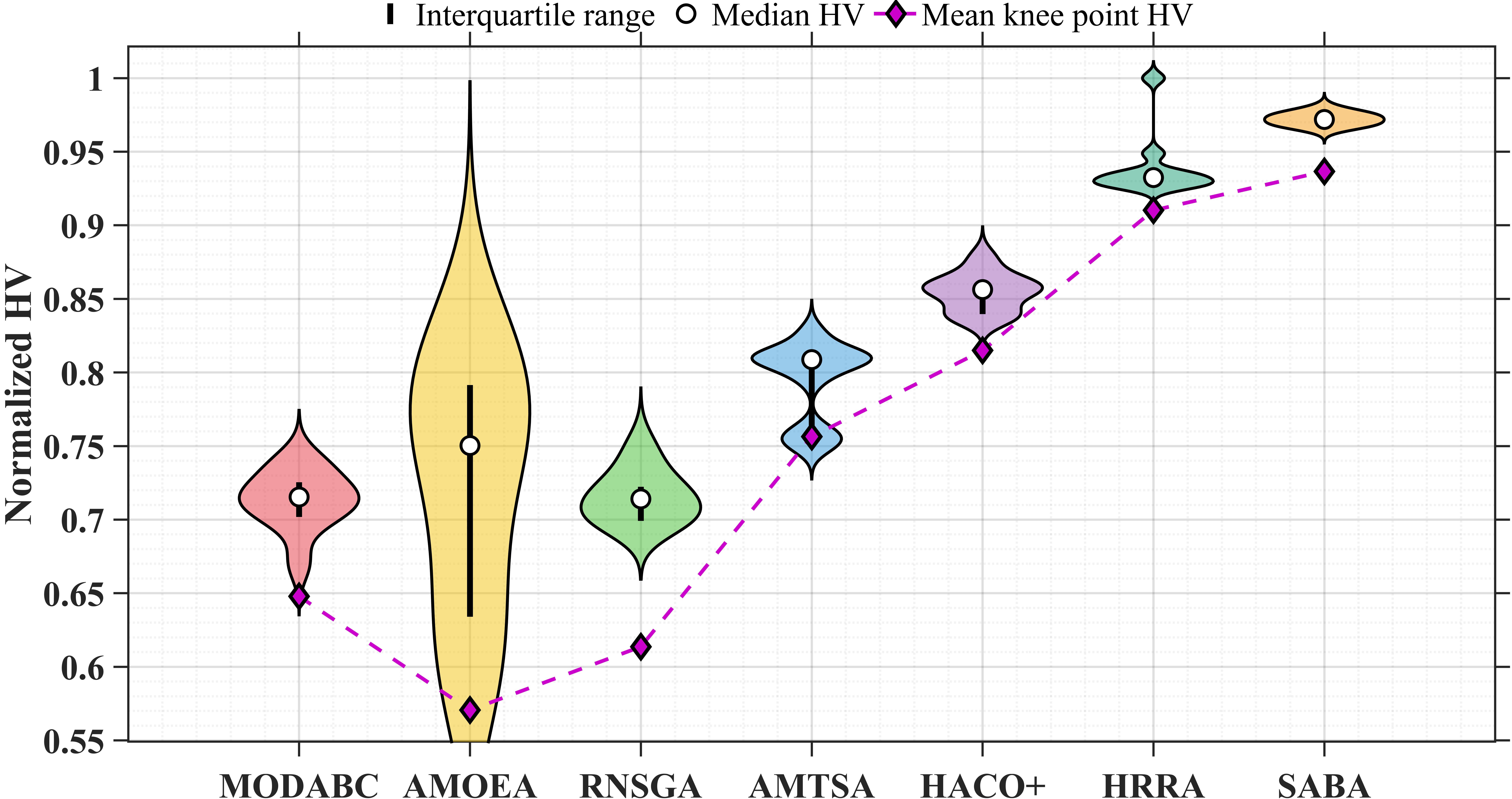}
    \vspace{3pt} % <-- 在这里插入这行代码
    \caption{Statistical results over ten independent runs}
    \label{fig:vio}
\end{figure}

To overcome the limitations and inherent stochasticity of a single execution, Fig.~\ref{fig:vio} presents a violin plot illustrating the statistical distribution of the HV indicator values obtained from ten independent runs on the instance to evaluate algorithms' performance consistency and robustness. The width of each violin's colored area reflects the probability density of the results (i.e., the frequency of a certain HV value), while its height represents the range of the distribution. The interquartile range (IQR) reflects the stability of the results (a shorter IQR indicates more stable performance), while the median HV reflects the algorithm's average performance level. Based on this, the results clearly reveal the comprehensive superiority of SABA. In terms of performance level, SABA's median HV (the white circle) is significantly higher than all competitors, reaffirming its ability to find superior-quality solution sets. More importantly, regarding robustness, its IQR bar is the shortest among all algorithms, and its entire violin shape is tightly concentrated at the top of the chart. This indicates that SABA's exceptional performance is not coincidental, as it can reliably converge to a near-optimal level with minimal variance across independent runs.

\subsection{Results and analysis on the benchmark suit}

Table~\ref{tab:results} summarizes the mean (standard deviation) of the HV indicator for each algorithm over 10 independent runs per instance, with the best mean value for each instance highlighted. For a rigorous statistical comparison, a Wilcoxon rank-sum test~\cite{chen2024archive} was conducted against SABA, where the symbols `$+$', `$-$', and `$\approx$' denote that the competitor's performance is significantly better than, worse than, or statistically equivalent to that of SABA, respectively.

\begin{figure}[htp]
    \centering
    \includegraphics[width=8.7cm]{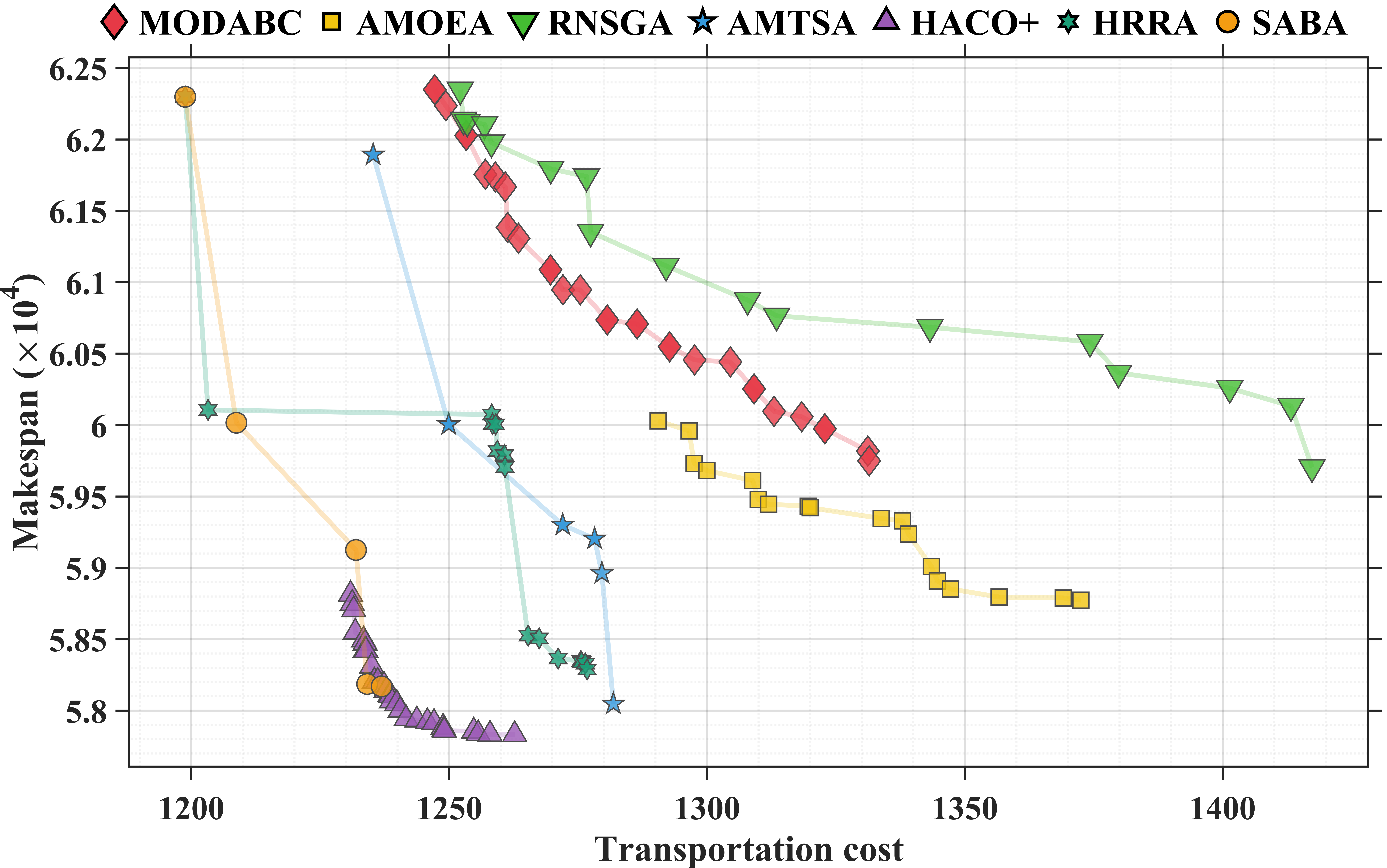}
    \caption{Algorithm performance comparison on Pro11}
    \label{fig:pro11}
\end{figure}

\begin{table*}[]
\centering
\caption{Comparison results on the proposed test set}
\label{tab:results}
\setlength{\tabcolsep}{8pt} % 列间距
% \centering % redundant with \centering outside table*
\small % 字体大小
\scalebox{0.65}{ % 缩放整个表格
\begin{tabular}{cccccccc}
\hline
\multicolumn{8}{c}{\textbf{Non-dominated solution set}} \\ % 表格子标题
\hline
Problem & MODABC~\cite{dai2023multi} & AMOEA~\cite{dong2024effective}   & RNSGA~\cite{10964773} & AMTSA~\cite{Chen2025adaptive} & HRRA~\cite{chen2025multiobjectivetaskallocationelectric} & HACO$+$~\cite{comert2023new} & SABA\\
\hline
Pro1    & 2.0944e-02 (8.12e-04) - & 2.1031e-02 (6.24e-03) - & 1.8706e-02 (2.85e-03) - & 2.0879e-02 (1.08e-03) - & 2.3978e-02 (7.32e-04) -                         & 2.4815e-02 (1.28e-03) -                         & \cellcolor{tablegray}9.0278e-02 (2.94e-03) \\
Pro2    & 4.1506e-02 (1.23e-03) - & 4.7933e-02 (9.55e-03) = & 4.4984e-02 (1.05e-02) = & 4.4517e-02 (1.27e-03) - & 5.3487e-02 (7.62e-04) =                         & \cellcolor{tablegray}5.6467e-02 (1.41e-03) + & 5.1057e-02 (7.96e-04)                         \\
Pro3    & 2.6859e-02 (9.29e-04) - & 2.9193e-02 (4.49e-03) - & 2.4607e-02 (1.31e-03) - & 3.0517e-02 (7.15e-04) - & 3.4815e-02 (7.99e-04) =                         & 3.3128e-02 (9.71e-04) -                         & \cellcolor{tablegray}3.5090e-02 (7.23e-04) \\
Pro4    & 3.4884e-02 (1.38e-03) - & 4.2658e-02 (5.95e-03) - & 3.4172e-02 (4.59e-03) - & 4.4073e-02 (1.02e-03) - & 5.2988e-02 (8.81e-04) -                         & 5.2046e-02 (5.93e-04) -                         & \cellcolor{tablegray}5.7427e-02 (1.78e-03) \\
Pro5    & 2.0846e-02 (6.59e-04) - & 2.4516e-02 (2.35e-03) - & 1.9949e-02 (5.43e-04) - & 2.1898e-02 (1.03e-03) - & 2.2883e-02 (4.47e-04) -                         & 2.8335e-02 (2.05e-04) -                         & \cellcolor{tablegray}2.2174e-02 (1.47e-04) \\
Pro6    & 2.7631e-02 (5.54e-04) - & 2.8351e-02 (5.14e-03) - & 3.0233e-02 (7.53e-03) - & 2.9894e-02 (4.73e-04) - & 3.1136e-02 (3.30e-03) -                         & 3.6661e-02 (3.21e-04) -                         & \cellcolor{tablegray}6.2256e-02 (5.38e-03) \\
Pro7    & 1.0911e-02 (5.88e-04) - & 1.1583e-02 (3.70e-03) - & 1.3721e-02 (8.50e-03) - & 1.2747e-02 (5.89e-04) - & 1.4082e-02 (5.00e-04) -                         & 1.4707e-02 (3.53e-04) -                         & \cellcolor{tablegray}1.5358e-02 (5.14e-04) \\
Pro8    & 1.9744e-02 (6.42e-04) - & 1.8571e-02 (2.17e-03) - & 1.9848e-02 (2.44e-03) - & 2.2206e-02 (6.40e-04) - & 2.5036e-02 (2.49e-04) -                         & 2.1787e-02 (2.61e-04) -                         & \cellcolor{tablegray}4.4664e-02 (3.51e-03) \\
Pro9    & 1.2819e-02 (3.60e-04) - & 1.3130e-02 (2.44e-03) - & 1.2652e-02 (9.18e-04) - & 1.3095e-02 (8.37e-04) - & 1.5470e-02 (2.69e-04) -                         & 1.7167e-02 (1.67e-04) -                         & \cellcolor{tablegray}1.7352e-02 (2.00e-04) \\
Pro10   & 1.9579e-02 (4.09e-04) - & 2.0410e-02 (1.57e-03) - & 1.8921e-02 (4.66e-04) - & 2.1182e-02 (4.72e-04) - & 2.3389e-02 (4.71e-04) -                         & 2.5266e-02 (1.28e-04) -                         & \cellcolor{tablegray}2.4492e-02 (2.10e-03) \\
Pro11   & 2.5731e-02 (7.98e-04) - & 2.3083e-02 (2.34e-03) - & 2.3924e-02 (1.45e-03) - & 3.0690e-02 (9.58e-04) - & 3.4870e-02 (3.29e-04) =                         & 3.3976e-02 (1.31e-04) -                         & \cellcolor{tablegray}3.4897e-02 (9.67e-04) \\
Pro12   & 1.4493e-02 (4.18e-04) - & 1.7085e-02 (2.63e-03) - & 1.4473e-02 (5.54e-04) - & 1.5333e-02 (4.05e-04) - & 1.8583e-02 (2.19e-04) -                         & 2.0880e-02 (2.24e-04) -                         & \cellcolor{tablegray}2.0148e-02 (3.08e-04) \\
Pro13   & 1.1399e-02 (3.62e-04) - & 1.0023e-02 (2.80e-03) - & 1.0977e-02 (4.99e-04) - & 1.3290e-02 (7.29e-04) - & 1.4409e-02 (1.27e-04) -                         & 1.6634e-02 (1.36e-04) -                         & \cellcolor{tablegray}1.6688e-02 (4.09e-04) \\
Pro14   & 1.0605e-02 (7.17e-04) - & 9.4229e-03 (1.68e-03) - & 1.0070e-02 (4.34e-04) - & 1.1425e-02 (6.84e-04) - & 1.4015e-02 (1.35e-04) -                         & 1.4601e-02 (1.48e-04) -                         & \cellcolor{tablegray}1.5319e-02 (3.38e-04) \\
Pro15   & 1.0377e-02 (3.71e-04) - & 9.6431e-03 (1.58e-03) - & 1.0129e-02 (3.96e-04) - & 1.1139e-02 (4.77e-04) - & 1.3674e-02 (3.17e-04) -                         & 1.3047e-02 (1.08e-04) -                         & \cellcolor{tablegray}1.5314e-02 (7.35e-06) \\
\hline
+/-/=   & 0/15/0                  & 0/14/1                  & 0/14/1                  & 0/15/0                  & 0/12/3                                          & 1/14/0                                          &                                               \\
\hline
\multicolumn{8}{c}{\textbf{Default output solution set}} \\ % 表格子标题
\hline
Problem & MODABC & AMOEA   & RNSGA & AMTSA & HRRA & HACO$+$ & SABA\\
\hline
Pro1    & 1.9766e-02 (1.19e-03) - & 1.7198e-02 (5.44e-03) - & 1.7302e-02 (8.23e-04) - & 2.0230e-02 (1.08e-03) - & 2.3528e-02 (1.84e-04) -                         & 2.3690e-02 (1.48e-03) -                         & \cellcolor{tablegray}8.1699e-02 (4.04e-03) \\
Pro2    & 3.1933e-02 (4.23e-03) - & 3.5600e-02 (1.21e-02) = & 3.1210e-02 (1.16e-02) - & 3.6609e-02 (3.73e-03) - & 4.5055e-02 (2.37e-03) =                         & \cellcolor{tablegray}5.4641e-02 (2.00e-03) + & 4.4356e-02 (9.34e-04)                         \\
Pro3    & 2.0923e-02 (3.08e-03) - & 2.4528e-02 (4.31e-03) - & 1.7650e-02 (4.14e-03) - & 2.9111e-02 (1.37e-03) = & \cellcolor{tablegray}3.1435e-02 (2.23e-03) = & 3.0995e-02 (1.44e-03) =                         & 3.0405e-02 (1.66e-03)                         \\
Pro4    & 2.2327e-02 (3.14e-03) - & 3.4601e-02 (1.01e-02) - & 2.2760e-02 (1.01e-02) - & 3.7966e-02 (5.54e-03) - & 5.0367e-02 (3.01e-03) =                         & 5.0889e-02 (8.65e-04) =                         & \cellcolor{tablegray}5.3008e-02 (3.55e-03) \\
Pro5    & 1.8920e-02 (1.04e-03) - & 2.0780e-02 (3.04e-03) - & 1.7389e-02 (1.67e-03) - & 1.9624e-02 (8.20e-04) - & 2.2804e-02 (4.32e-04) -                         & 2.5513e-02 (1.10e-03) -                         & \cellcolor{tablegray}2.7691e-02 (3.77e-04) \\
Pro6    & 2.1438e-02 (1.90e-03) - & 2.0561e-02 (4.80e-03) - & 2.3680e-02 (1.03e-02) - & 2.3957e-02 (1.78e-03) - & 2.6022e-02 (1.15e-03) -                         & 3.4615e-02 (5.57e-04) -                         & \cellcolor{tablegray}6.1332e-02 (2.11e-03) \\
Pro7    & 1.0062e-02 (6.50e-04) - & 9.9311e-03 (3.45e-03) - & 1.2695e-02 (8.92e-03) - & 1.2243e-02 (4.28e-04) - & 1.3332e-02 (7.10e-04) -                         & 1.2979e-02 (7.89e-04) -                         & \cellcolor{tablegray}1.5153e-02 (2.51e-04) \\
Pro8    & 1.6950e-02 (1.16e-03) - & 1.5063e-02 (3.35e-03) - & 1.6299e-02 (4.39e-03) - & 2.1647e-02 (7.80e-04) - & 2.1055e-02 (4.69e-03) -                         & 1.9158e-02 (7.32e-04) -                         & \cellcolor{tablegray}4.2951e-02 (2.44e-03) \\
Pro9    & 1.1304e-02 (7.71e-04) - & 1.1369e-02 (3.05e-03) - & 1.1504e-02 (1.29e-03) - & 1.2162e-02 (1.32e-03) - & 1.5116e-02 (7.02e-04) -                         & 1.5833e-02 (3.41e-04) -                         & \cellcolor{tablegray}1.7444e-02 (2.31e-04) \\
Pro10   & 1.5974e-02 (9.32e-04) - & 1.8440e-02 (2.35e-03) - & 1.5408e-02 (9.95e-04) - & 1.9436e-02 (1.76e-03) - & 2.2510e-02 (1.13e-03) -                         & 2.3474e-02 (7.13e-04) -                         & \cellcolor{tablegray}2.5067e-02 (1.32e-03) \\
Pro11   & 2.0790e-02 (7.70e-04) - & 1.9327e-02 (3.57e-03) - & 1.9536e-02 (1.46e-03) - & 2.3714e-02 (3.30e-03) - & 2.8098e-02 (1.69e-03) =                         & \cellcolor{tablegray}3.2910e-02 (2.97e-04) + & 2.9376e-02 (1.03e-03)                         \\
Pro12   & 1.2658e-02 (6.83e-04) - & 1.5713e-02 (2.97e-03) - & 1.2644e-02 (6.08e-04) - & 1.4801e-02 (6.42e-04) - & 1.7485e-02 (7.00e-04) -                         & 1.8589e-02 (6.11e-04) -                         & \cellcolor{tablegray}2.0628e-02 (7.16e-04) \\
Pro13   & 1.0920e-02 (3.98e-04) - & 8.7361e-03 (2.63e-03) - & 1.0472e-02 (6.50e-04) - & 1.2985e-02 (1.09e-03) - & 1.4004e-02 (1.09e-03) -                         & 1.5860e-02 (4.21e-04) -                         & \cellcolor{tablegray}1.6901e-02 (3.66e-04) \\
Pro14   & 1.0118e-02 (6.69e-04) - & 8.0884e-03 (2.29e-03) - & 9.3623e-03 (5.53e-04) - & 1.1179e-02 (5.39e-04) - & 1.2931e-02 (4.13e-04) -                         & 1.3946e-02 (4.16e-04) -                         & \cellcolor{tablegray}1.5093e-02 (2.51e-04) \\
Pro15   & 9.7176e-03 (3.33e-04) - & 8.7151e-03 (2.13e-03) - & 9.3951e-03 (6.03e-04) - & 1.0843e-02 (6.16e-04) - & 1.3312e-02 (4.99e-04) -                         & 1.2451e-02 (3.04e-04) -                         & \cellcolor{tablegray}1.4085e-02 (6.62e-04) \\
\hline
+/-/=   & 0/15/0                  & 0/14/1                  & 0/15/0                  & 0/14/1                  & 0/11/4                                          & 2/11/2                                          &                                              
\\
\hline
\end{tabular}
} % end scalebox
\end{table*}

Analyzing the overall results, SABA demonstrates superior performance on nearly all instances, with the sole exception of Pro2. As defined in Section~\ref{s:problem}, Pro2 features the smallest task scale and, consequently, the shortest algorithm runtime. Under such tight runtime constraints, the powerful initialization of HACO$^+$ provides it with a significant head-start advantage that strongly influences the final outcome. This observation corroborates our analysis of HACO$^+$'s initialization capabilities in Section~\ref{s:realworld}. Additionally, this instance features a high robot-to-task ratio, which increases the computational effort required by SABA's complex balancing mechanisms. On such a small-scale problem with a very short runtime, this overhead can marginally outweigh the benefits of its structural optimization. However, its dominant performance across the entire benchmark suite decisively proves SABA's robustness, scalability, and unparalleled capability in handling complex, large-scale problems.

\begin{figure}[htp]
    \centering
    \includegraphics[width=8.7cm]{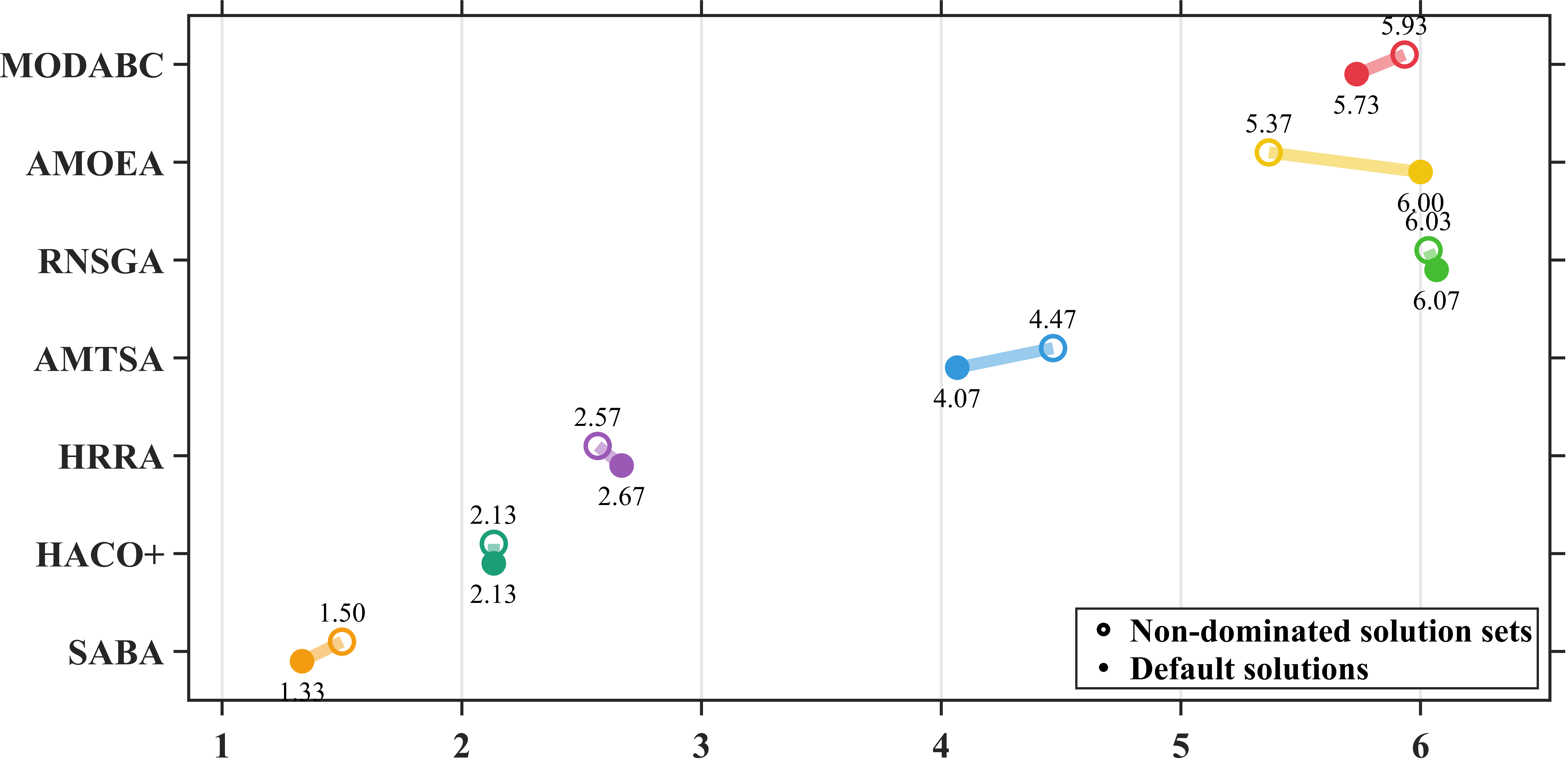}
    \caption{Statistical comparison of algorithm Friedman rankings}
    \label{fig:fri}
\end{figure}

Regarding the default compromise solutions, besides the aforementioned Pro2 case, HACO$^+$ also achieved a better knee point solution on instance Pro11. This can also be attributed to its powerful initialization stage, which occasionally locates a high-quality solution region early in the search. However, its weaker capability in maintaining diversity leads to a final solution set that is highly clustered around that point. As illustrated in Fig.~\ref{fig:pro11}, the entire PF of HACO$^+$ occupies a narrow subspace adjacent to one of the extreme solutions on SABA's more diverse front. This phenomenon confirms the performance instability of HACO$^+$ mentioned in Section~\ref{s:realworld}. In contrast, SABA, in its pursuit of global solution quality, produces a PF with a wide span across both objectives. This provides decision-makers with a richer portfolio of choices to accommodate diverse time-cost preferences. Therefore, compared to the extremeness of HACO$^+$'s solution, SABA's default output represents a better trade-off between the two objectives on this problem. Considering the entire test set, SABA also possesses better generality, which is consistent with our analysis of the overall solution sets.

To provide a rigorous statistical evaluation of the overall algorithm performance, Fig.~\ref{fig:fri} presents the average ranks obtained from a Friedman test conducted on the HV indicators. The test results clearly indicate that SABA achieved the lowest (i.e., best) average rank, with a statistically significant difference from its peers. This finding further corroborates the conclusions drawn in the previous analyses.

\section{Conclusion and future works}

This study addressed the complex agricultural multi-robot task allocation problem, incorporating the coupled constraints of task splitting, payload, and battery capacity. To tackle this challenge, we proposed a segment anchoring-based balancing algorithm (SABA), which aims to co-optimize system makespan and total transportation cost. SABA's novelty lies in its two synergistic mechanisms: the sequential anchoring and balancing mechanism (SABM), which introduces an anchor-and-reconstruct paradigm to systematically handle disruptions from energy constraints, and the proportional splitting-based rebalancing mechanism (PSRM) for fine-grained makespan tuning.

Extensive experiments against six state-of-the-art algorithms on a real-world case study and a diverse benchmark suite validated SABA's superiority. The results consistently demonstrated that SABA provides a richer and more effective portfolio of scheduling options, achieving a dominant advantage in both solution quality and robustness. Theoretically, this research contributes a new framework for a class of complex routing problems involving dynamic interruptions. Practically, SABA serves as an efficient and reliable decision-support tool for automated systems in smart agriculture and logistics.

For future work, this research can be further extended by incorporating more realistic characteristics. Primary avenues for exploration include adapting the SABA framework for dynamic and stochastic environments requiring real-time re-planning, and extending it to address the collaborative optimization of heterogeneous robot fleets with diverse capabilities. Furthermore, integrating more sophisticated battery health models into the objectives is also a worthy topic for future investigation.

%\vspace{-1em}
\bibliographystyle{unsrt}
\bibliography{ref}

\begin{thebibliography}{10}

\bibitem{chen2024efficient}
Tinghui Chen, Weiyi Yang, Zhetao Zhang, and Xin Luo.
\newblock An efficient industrial robot calibrator with multiplaner constraints.
\newblock {\em IEEE Transactions on Industrial Informatics}, 2024.

\bibitem{asmara2024economic}
Alla Asmara, Agung~Satria Permana, and Mohammad~Iqbal Irfany.
\newblock Economic determinants of changes in labor participation in the indonesian agricultural sector before and during covid-19.
\newblock {\em Journal of the International Society for Southeast Asian Agricultural Sciences}, 30(1):62--79, 2024.

\bibitem{ranjha2021facilitating}
Ali Ranjha, Georges Kaddoum, and Kapal Dev.
\newblock Facilitating urllc in uav-assisted relay systems with multiple-mobile robots for 6g networks: A prospective of agriculture 4.0.
\newblock {\em IEEE Transactions on Industrial Informatics}, 18(7):4954--4965, 2021.

\bibitem{ka2024systematic}
Athira KA and Umashankar Subramaniam.
\newblock A systematic literature review on multi-robot task allocation.
\newblock {\em ACM Computing Surveys}, 57(3):1--28, 2024.

\bibitem{dong2024effective}
Jin-Shuai Dong, Quan-Ke Pan, Zhong-Hua Miao, Hong-Yan Sang, and Liang Gao.
\newblock An effective multi-objective evolutionary algorithm for multiple spraying robots task assignment problem.
\newblock {\em Swarm and Evolutionary Computation}, 87:101558, 2024.

\bibitem{dai2023multi}
Loulei Dai, Quanke Pan, Zhonghua Miao, Ponnuthurai~Nagaratnam Suganthan, and Kaizhou Gao.
\newblock Multi-objective multi-picking-robot task allocation: mathematical model and discrete artificial bee colony algorithm.
\newblock {\em IEEE Transactions on Intelligent Transportation Systems}, (6):6061--6073, 2023.

\bibitem{guo2024effective}
Hengwei Guo, Zhonghua Miao, JC~Ji, and Quanke Pan.
\newblock An effective collaboration evolutionary algorithm for multi-robot task allocation and scheduling in a smart farm.
\newblock {\em Knowledge-Based Systems}, 289:111474, 2024.

\bibitem{wang2024multi}
Cun-Hai Wang, Quan-Ke Pan, Xiao-Ping Li, Hong-Yan Sang, and Bing Wang.
\newblock A multi-objective teaching-learning-based optimizer for a cooperative task allocation problem of weeding robots and spraying drones.
\newblock {\em Swarm and Evolutionary Computation}, 87:101565, 2024.

\bibitem{guo2024hybrid}
Xiang Guo, Zhong-Hua Miao, Quan-Ke Pan, and Xuan He.
\newblock Hybrid loading situation vehicle routing problem in the context of agricultural harvesting: A reconstructed moea/d with parallel populations.
\newblock {\em Swarm and Evolutionary Computation}, 91:101730, 2024.

\bibitem{Chen2025adaptive}
Peng Chen, Jing Liang, Kang-Jia Qiao, Hui Song, Cai-Tong Yue, Kun-Jie Yu, Ponnuthurai~Nagaratnam Suganthan, and Witold Pedrycz.
\newblock {An adaptive multi-objective task splitting algorithm for agricultural multi-robot task allocation}.
\newblock \url{https://anonymous.4open.science/r/AMTSA}, 2025.
\newblock Preprint, under review at a peer-reviewed journal.

\bibitem{bruglieri2023matheuristic}
Maurizio Bruglieri, Massimo Paolucci, and Ornella Pisacane.
\newblock A matheuristic for the electric vehicle routing problem with time windows and a realistic energy consumption model.
\newblock {\em Computers \& Operations Research}, 157:106261, 2023.

\bibitem{bavand2022online}
Amin Bavand, S~Ali Khajehoddin, Masoud Ardakani, and Ahmadreza Tabesh.
\newblock Online estimations of li-ion battery soc and soh applicable to partial charge/discharge.
\newblock {\em IEEE Transactions on transportation electrification}, 8(3):3673--3685, 2022.

\bibitem{tarar2023techno}
Muhammad~Osama Tarar, Naveed~UL Hassan, Ijaz~Haider Naqvi, and Michael Pecht.
\newblock Techno-economic framework for electric vehicle battery swapping stations.
\newblock {\em IEEE Transactions on Transportation Electrification}, 9(3):4458--4473, 2023.

\bibitem{chen2025multiobjectivetaskallocationelectric}
Peng Chen, Jing Liang, Hui Song, Kang-Jia Qiao, Cai-Tong Yue, Kun-Jie Yu, Ponnuthurai~Nagaratnam Suganthan, and Witold Pedrycz.
\newblock {Multi-objective task allocation for electric harvesting robots: a hierarchical route reconstruction approach}.
\newblock {\em arXiv preprint arXiv:2509.11025}, 2025.
\newblock Preprint, under review at a peer-reviewed journal.

\bibitem{yilmaz2022variable}
Yusuf Yilmaz and Can~B Kalayci.
\newblock Variable neighborhood search algorithms to solve the electric vehicle routing problem with simultaneous pickup and delivery.
\newblock {\em Mathematics}, 10(17):3108, 2022.

\bibitem{guo2023low}
Yinan Guo, Yao Huang, Shirong Ge, Yizhe Zhang, Ersong Jiang, Bin Cheng, and Shengxiang Yang.
\newblock Low-carbon routing based on improved artificial bee colony algorithm for electric trackless rubber-tyred vehicles.
\newblock {\em Complex System Modeling and Simulation}, 3(3):169--190, 2023.

\bibitem{fan2024two}
Lijun Fan.
\newblock A two-stage hybrid ant colony algorithm for multi-depot half-open time-dependent electric vehicle routing problem.
\newblock {\em Complex \& Intelligent Systems}, 10(2):2107--2128, 2024.

\bibitem{zhou2022multi}
Binghai Zhou and Zhe Zhao.
\newblock Multi-objective optimization of electric vehicle routing problem with battery swap and mixed time windows.
\newblock {\em Neural Computing and Applications}, 34(10):7325--7348, 2022.

\bibitem{comert2023new}
Serap~Ercan Comert and Harun~Resit Yazgan.
\newblock A new approach based on hybrid ant colony optimization-artificial bee colony algorithm for multi-objective electric vehicle routing problems.
\newblock {\em Engineering Applications of Artificial Intelligence}, 123:106375, 2023.

\bibitem{choudhury2022dynamic}
Shushman Choudhury, Jayesh~K Gupta, Mykel~J Kochenderfer, Dorsa Sadigh, and Jeannette Bohg.
\newblock Dynamic multi-robot task allocation under uncertainty and temporal constraints.
\newblock {\em Autonomous Robots}, 46(1):231--247, 2022.

\bibitem{xiong2019development}
Ya~Xiong, Cheng Peng, Lars Grimstad, P{\aa}l~Johan From, and Volkan Isler.
\newblock Development and field evaluation of a strawberry harvesting robot with a cable-driven gripper.
\newblock {\em Computers and electronics in agriculture}, 157:392--402, 2019.

\bibitem{eckert2012a+}
Kerstin Eckert, Laurence Rongy, and Anne De~Wit.
\newblock A+ b→ c reaction fronts in hele-shaw cells under modulated gravitational acceleration.
\newblock {\em Physical Chemistry Chemical Physics}, 14(20):7337--7345, 2012.

\bibitem{valero2017influence}
Francisco Valero, Francisco Rubio, Carlos Llopis-Albert, and Juan~Ignacio Cuadrado.
\newblock Influence of the friction coefficient on the trajectory performance for a car-like robot.
\newblock {\em Mathematical Problems in Engineering}, 2017(1):4562647, 2017.

\bibitem{DOE2024EV}
{U.S. Department of Energy}.
\newblock All-electric vehicles.
\newblock \url{https://www.fueleconomy.gov/feg/evtech.shtml}, 2024.
\newblock Accessed: December 12, 2024.

\bibitem{mcnulty2022review}
David McNulty, Aaron Hennessy, Mei Li, Eddie Armstrong, and Kevin~M Ryan.
\newblock A review of li-ion batteries for autonomous mobile robots: Perspectives and outlook for the future.
\newblock {\em Journal of Power Sources}, 545:231943, 2022.

\bibitem{mitici2022electric}
Mihaela Mitici, Madalena Pereira, and Fabrizio Oliviero.
\newblock Electric flight scheduling with battery-charging and battery-swapping opportunities.
\newblock {\em EURO Journal on Transportation and Logistics}, 11:100074, 2022.

\bibitem{yang2015battery}
Jun Yang and Hao Sun.
\newblock Battery swap station location-routing problem with capacitated electric vehicles.
\newblock {\em Computers \& operations research}, 55:217--232, 2015.

\bibitem{lou2024analysis}
Kairan Lou, Zongbin Wang, Bin Zhang, Qiu Xu, Wei Fu, Yang Gu, and Jinyi Liu.
\newblock Analysis and experimentation on the motion characteristics of a dragon fruit picking robot manipulator.
\newblock {\em Agriculture}, 14(11):2095, 2024.

\bibitem{miller1960integer}
Clair~E Miller, Albert~W Tucker, and Richard~A Zemlin.
\newblock Integer programming formulation of traveling salesman problems.
\newblock {\em Journal of the ACM (JACM)}, 7(4):326--329, 1960.

\bibitem{10964773}
Peng Chen, Jing Liang, Kang-Jia Qiao, Hui Song, Ponnuthurai~Nagaratnam Suganthan, Lou-Lei Dai, and Xuan-Xuan Ban.
\newblock A reinforced neighborhood search method combined with genetic algorithm for multi-objective multi-robot transportation system.
\newblock {\em IEEE Transactions on Intelligent Transportation Systems}, pages 1--14, 2025.

\bibitem{englert2014worst}
Matthias Englert, Heiko R{\"o}glin, and Berthold V{\"o}cking.
\newblock Worst case and probabilistic analysis of the 2-opt algorithm for the tsp.
\newblock {\em Algorithmica}, 68(1):190--264, 2014.

\bibitem{chen2024archive}
Peng Chen, Zhimeng Li, Kangjia Qiao, PN~Suganthan, Xuanxuan Ban, Kunjie Yu, Caitong Yue, and Jing Liang.
\newblock An archive-assisted multi-modal multi-objective evolutionary algorithm.
\newblock {\em Swarm and Evolutionary Computation}, 91:101738, 2024.

\bibitem{Kaplan2025Knee}
Dmitry Kaplan.
\newblock Knee point.
\newblock \url{https://www.mathworks.com/matlabcentral/fileexchange/35094-knee-point}, 2025.
\newblock MATLAB Central File Exchange. Accessed: 2025-04-17.

\end{thebibliography}
\end{document}

\section*{Supplementary materials}

% \begin{center}
% \begin{minipage}{0.4\textwidth} % 可以调整0.7这个比例来控制文本框的宽度
% \setlength{\itemsep}{6pt}
% The supplementary materials for HRRA includes:
% \begin{itemize}
%     \item Section \ref{S-complexity}: algorithm complexity analysis;

%     \item Section \ref{S-parameter}: parameter sensitivity analysis;

%     \item Section \ref{S-ablation}: ablation study;

%     \item Section \ref{S-default outputs}: performance comparison of the default outputs;

%     \item Section \ref{S-Algorithms}: presentation of Algorithms~\ref{alg:initialization}~-~\ref{alg:SRRM};

%     \item Section \ref{S-table}: detailed results in Tables~\ref{resultsr=4}~-~\ref{resultsr=6};

%     \item Section \ref{S-figures}: detailed results in Figs.~\ref{fig:r4_1}~-~\ref{fig:r6_1}.
% \end{itemize}
% \end{minipage}
% \end{center}

\section{Computational complexity}

The computational complexity of the proposed algorithm is analyzed as follows. The main loop of the algorithm is derived from our prior work, AMTSA, and has a per-iteration complexity of $O(n \cdot k)$~\cite{Chen2025adaptive}, where $n$ is the total number of tasks and $k$ is the average number of tasks per cycle. The core complexity of the proposed SABA algorithm stems from the probabilistic invocation of the SABM and the final execution of the PSRM.

Within SABM, the CSOS adjusts the execution sequence of cycles for all robots. This is achieved by a 2-opt search over an average of $c$ cycles for each of the $r$ robots, where each swap requires a path evaluation of $O(n/r)$. This results in a total complexity of approximately $O(r \cdot c^2 \cdot (n/r)) \approx O(c^2 \cdot n)$. The SAS iteratively initializes the remaining tasks, with a complexity of approximately $O(c \cdot k)$~\cite{Chen2025adaptive} for the initialization of each sub-problem. Assuming a worst-case scenario where a recharge is needed after every cycle, the SAS iterates $c$ times for each robot. Given that SAS invokes CSOS in each iteration, the complexity of the SAS is approximately $O(c^2 \cdot k \cdot r + c^3 \cdot n) \approx O(c^3 \cdot n)$. The final RWBS needs to re-plan the entire pool of residual tasks (approaching $n$ in the worst case), resulting in a complexity of about $O(n^2)$. In contrast, the PSRM has a much lower complexity; its dominant operation is the evaluation of all paths to identify the bottleneck, with a complexity of approximately $O(n)$.

In summary, the cost of a single SABM execution is dominated by the SAS, with a complexity of $O(c^3 \cdot n)$. Given that SABM is a computationally intensive operation, SABA employs the parameter $p$ to control its invocation probability, thereby striking a balance between solution quality and computational efficiency.

\section{Parameter sensitivity and component effectiveness analysis}

To determine the optimal configuration of its core parameters and to validate the effectiveness of the proposed key components, this section presents a parameter sensitivity analysis and a component ablation study. For this purpose, we designed several variants of the SABA algorithm. To analyze the sensitivity of the parameter controlling the invocation frequency of SABM, we configured five variants, P1--P5, where the invocation probability $p$ was uniformly set to values in $\{0.2, 0.4, 0.6, 0.8, 1.0\}$, respectively. Furthermore, setting the parameter $p$ to 0 effectively deactivates the SABM throughout the entire search process. This variant, denoted as V1, thus serves to verify the impact of SABM on SABA. Correspondingly, variant V2 removes the PSRM, and variant V3 removes both mechanisms. All variants were executed for 10 independent runs on each test instance. We employ the win-count as the core evaluation metric herein: in a single run, a variant is credited with a win if the HV indicator of its final non-dominated set is superior to that of all its competitors.

\begin{figure}[htp]
    \centering
    \includegraphics[width=15cm]{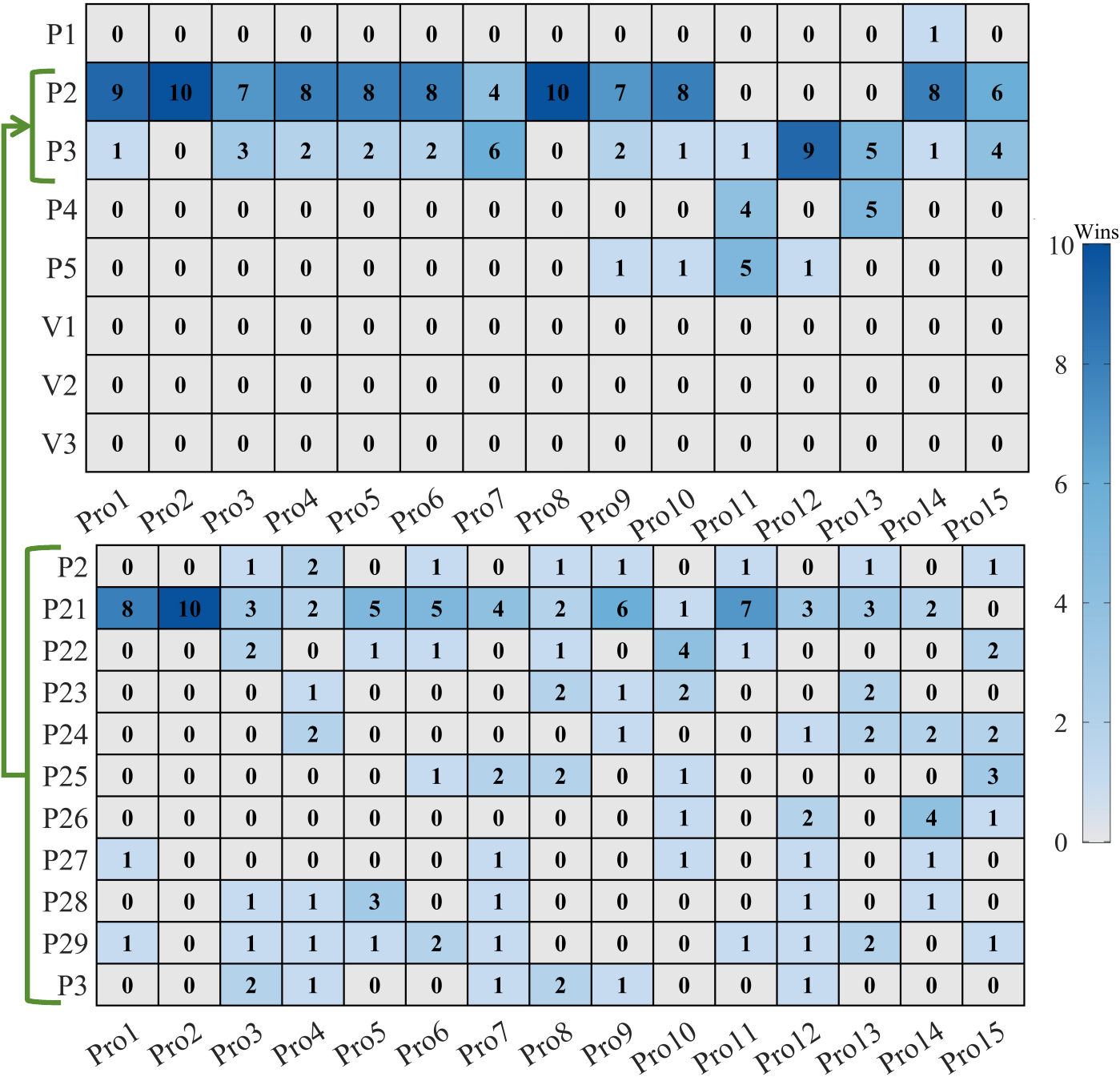}
    \caption{Results of parameter sensitivity and component effectiveness analysis}
    \label{fig:ablation_param}
\end{figure}

The win-count statistics, summarized in Fig.~\ref{fig:ablation_param}, clearly reveal that the SABM invocation probability $p$ is a critical parameter that governs the algorithm's core behavior, directly regulating the trade-off between global exploration and local exploitation. A darker shade in a square indicates a higher number of wins. Configurations with either very low ($p \le 0.2$) or very high ($p \ge 0.8$) values exhibited suboptimal performance. This indicates that an infrequent invocation fails to fully leverage SABM's powerful solution-reconstruction capabilities to escape local optima, whereas an overly frequent invocation consumes excessive computational resources and diminishes population diversity, leading to premature convergence. In contrast, the variants with moderate frequencies (0.4 and 0.6) performed exceptionally well, demonstrating their effectiveness in striking this exploration-exploitation balance. A noteworthy phenomenon is that on mega-scale instances such as Pro11--13, where both the number of tasks $n$ and robots $r$ are at extreme levels, the performance of P2 ($p=0.4$) was inferior to that of P3 ($p=0.6$). This suggests that when confronted with an exceptionally large and complex search space, more frequent and aggressive intervention from SABM becomes critical for breaking search stagnation. To explore the optimal balance point between these two variants, we further designed variants P21--P29, which uniformly fill the parameter space in the interval $[0.4, 0.6]$ with an increment of 0.02. The final experimental results indicated that while several parameter values showed strong performance on specific instances, P21 ($p=0.42$) demonstrated the best robustness in terms of its consistent, high-level performance across the entire benchmark suite. Consequently, $p=0.42$ was selected as the definitive parameter configuration for the SABA algorithm in all subsequent comparative experiments.

The fact that variants V1-V3 failed to achieve the optimal performance metric in any test instance clearly confirms the necessity and synergistic effect of our two proposed mechanisms. SABM serves as the core engine for constructing a high-quality solution structure, providing a solid foundation for subsequent optimization, while PSRM acts as the critical final step that performs fine-grained balancing on this foundation to ensure the solution's superiority.

\clearpage

\section{Algorithms}
\label{S-Algorithms}

\begin{algorithm}[htb]
\SetAlgoLined
\KwIn{%
    \begin{tabular}[t]{ll}
        A non-dominated solution: & $solution$\\
    \end{tabular}
}
\KwOut{An improved solution: $new\_solution$}
%\vspace{0.1em}
$new\_solution \leftarrow solution$ \tcp{Initialize with the input solution}
%\vspace{0.1em}
\tcp{Per-robot optimization (CSOS and SAS)}
\ForEach{$robot_k \in new\_solution$}{
    Get the full path $path_k$ for robot $k$\;
    \tcp{CSOS: Optimize the macro-level sequence}
    Decompose $path_k$ into a set of cycles $S_k$\;
    $S'_k$ \leftarrow $S_k$ \tcp{Sequence order with 2-opt}
    Rebuild path $path'$ from the ordered cycles in $S'_k$\;
    Apply 2-opt to tasks within the first cycle of $path'$ disrupted by a charge stop\;
    %\vspace{0.1em}
    \tcp{SAS: Iteratively anchor and reconstruct}
    $final\_anchored\_path \leftarrow \emptyset$\;
    $current\_path\_segment \leftarrow path'$\;
    \While{\text{true}}{
        Find the index of the first charge stop $(-1)$ in $current\_path\_segment$ as $anchor\_idx$\;
        \If{$anchor\_idx$ is null}{
            Append $current\_path\_segment$ to $final\_anchored\_path$\;
            \textbf{break}\;
        }
        $newly\_anchored \leftarrow$ the segment of $current\_path\_segment$ from start to $anchor\_idx$\;
        Append $newly\_anchored$ to $final\_anchored\_path$\;
        Extract all tasks from the segment after $anchor\_idx$ into $residual\_tasks$\;
        \If{$residual\_tasks$ is empty}{
            \textbf{break}\;
        }
        Generate a new path segment for $residual\_tasks$ \tcp{Reinitialization}
        $current\_path\_segment \leftarrow$ the newly generated path segment\;
    }
    Update path of $robot_k$ in $new\_solution$ with $final\_anchored\_path$\;
}
\vspace{0.1em}
\tcp{Global balancing (RWBS)}
$residual\_pool \leftarrow \emptyset$\;
$fixed\_makespans \leftarrow \text{an array of zeros for } r \text{ robots}$\;
\ForEach{$robot_k \in new\_solution$}{
    Split the path of $robot_k$ at its last charge stop into a $fixed\_part$ and a $residual\_part$\;
    Add tasks from $residual\_part$ to the $residual\_pool$\;
    Calculate the makespan of $fixed\_part$ and store it in $fixed\_makespans[k]$\;
    Update the path of $robot_k$ in $new\_solution$ to be its $fixed\_part$\;
}
Aggregate demands for all tasks in the $residual\_pool$\;
Re-plan tasks in $residual\_pool$ into a new set of optimized cycles $S_{\text{new}}$\;
Calculate the duration for each cycle in $S_{\text{new}}$ \tcp{Reinitialization}
Assign cycles from $S_{\text{new}}$ to robots greedily based on $fixed\_makespans$ to balance workload\;
Update $new\_solution$ by appending the assigned new cycles to each robot's path\;
%\vspace{0.1em}
\Return{$new\_solution$}
\caption{Sequential anchoring and balancing mechanism (SABM)}
\label{alg:sabm}
\end{algorithm}

\begin{algorithm}[htb]
\SetAlgoLined
\KwIn{%
    \begin{tabular}[t]{ll}
        A non-dominated solution: & $solution$\\
    \end{tabular}
}
\KwOut{A makespan-balanced solution: $new\_solution$}
%\vspace{0.1em}
$new\_solution \leftarrow solution$ \tcp{Initialize with the input solution}
%\vspace{0.1em}
\tcp{Identify the bottleneck and donor cycle}
Calculate makespans for all robots in $solution$\;
Identify the bottleneck robot $r_b$ with the maximum makespan $T_{\text{max}}$\;
Find the last charge stop in the path of $r_b$\;
Decompose the path of $r_b$ after the last charge stop into a set of cycles $S_{\text{after}}$\;
\If{$S_{\text{after}}$ is empty}{
    \Return{$solution$} \tcp{No cycles available to split}
}
Select the cycle $s_{\text{donor}}$ from $S_{\text{after}}$ with the lowest transportation expense\;
Remove $s_{\text{donor}}$ from the path of $r_b$ in $new\_solution$\;

%\vspace{0.1em}
\tcp{Calculate baseline times and split ratios}
Re-evaluate and store the makespans of all robots as $baseline\_makespans$\;
Calculate the total available time $T_{\text{total}}$ by summing $baseline\_makespans$ and the duration of $s_{\text{donor}}$\;
Calculate the ideal balanced makespan $T_{\text{ideal}} \leftarrow \frac{T_{\text{total}}}{r}$\;
\ForEach{robot $k \in R$}{
    Calculate the time gap for robot $k$: $\Delta T_k \leftarrow T_{\text{ideal}} - baseline\_makespans[k]$\;
}
Calculate the proportional split ratios for $s_{\text{donor}}$ based on all positive time gaps $\Delta T_k$\;

%\vspace{0.1em}
\tcp{Reallocate split portions of the donor cycle}
\ForEach{robot $k \in R$}{
    \If{robot $k$ has a positive split ratio}{
        Create a new cycle $s_{\text{new}}$ for robot $k$ with the same tasks as $s_{\text{donor}}$\;
        Set the demand completion ratio for tasks in $s_{\text{new}}$ according to the calculated split ratio\;
        Append $s_{\text{new}}$ to the path of robot $k$ in $new\_solution$\;
        Update the corresponding splitting information\;
    }
}
%\vspace{0.1em}
\Return{$new\_solution$}
\caption{Proportional splitting-based rebalancing mechanism (PSRM)}
\label{alg:psrm}
\end{algorithm}

\clearpage

\bibliographystyle{unsrt}
\bibliography{ref}